\documentclass[%
twocolumn,
superscriptaddress,
 amsmath,amssymb,
 aps,prx
]{revtex4-2}

\usepackage{bbold}
\usepackage{amsfonts}
\usepackage{dsfont}
\usepackage{xcolor}
\usepackage{bm}
\usepackage{hyperref}
\hypersetup{plainpages=false,colorlinks=true,linkcolor=blue, citecolor=blue, urlcolor=blue}

\usepackage{xcolor}
\usepackage{bbold}
\usepackage{mathrsfs}
\usepackage{natbib}
\usepackage{here}

\usepackage{graphicx}

\usepackage{array,mathtools,amssymb,booktabs}
\newcolumntype{C}{>{$}c<{$}}

\begin{document}
\title{Hidden Markov model analysis for fluorescent time series of quantum dots}
\author{Tatsuhiro Furuta}\email{furuta.tatsuhiro704@mail.kyutech.jp}
\affiliation{Graduate School of Engineering, Kyushu Institute of Technology, Kitakyushu 804-8550, Japan}
\author{Keisuke Hamada}
\affiliation{Graduate School of Engineering, Kyushu Institute of Technology, Kitakyushu 804-8550, Japan}
\author{Masaru Oda}
\affiliation{Graduate School of Engineering, Kyushu Institute of Technology, Kitakyushu 804-8550, Japan}
\author{Kazuma Nakamura}\email{kazuma@mns.kyutech.ac.jp}
\affiliation{Graduate School of Engineering, Kyushu Institute of Technology, Kitakyushu 804-8550, Japan}

\begin{abstract}
We present a hidden Markov model analysis for fluorescent time series of quantum dots. A fundamental quantity to measure optical performance of the quantum dots is a distribution function for the light-emission duration. So far, to estimate it, a threshold value for the fluorescent intensity was introduced, and the light-emission state was evaluated as a state above the threshold. With this definition, the light-emission duration was estimated, and its distribution function was derived as a blinking plot. Due to the noise in the fluorescent data, however, this treatment generates a large number of artificially short-lived emission states, thus leading to an erroneous blinking plot. In the present paper, we propose a hidden Markov model to eliminate these artifacts. The hidden Markov model introduces a hidden variable specifying the light-emission and quenching states behind the observed fluorescence. We found that it is possible to avoid the above artifacts by identifying the state from the hidden-variable time series. We found that, from the analysis of experimental and theoretical benchmark data, the accuracy of our hidden Markov model is beyond human cognitive ability.
\end{abstract} 

\maketitle

\section{Introduction}\label{sec_Introduction}
Recently, studies on analyzing various material data by machine learning technique to propose new materials and new synthetic methods have become active~\cite{Seko_2014, Seko_2015, Lee_2016, Osada_2020, Severson_2019, Jiang_1998, Stanev_2021, Tao_2021, Peterson_2021}. The development of databases including material structures and their physical properties is progressing~\cite{Jain_2013,Ong_2013,Ong_2015,Jong_2015}, and the development of software for facilitating machine-learning analyses is also making great progress~\cite{Pedregosa_2011,Chang_2011}. Most of the research so far has focused on the correlation analysis of static data of materials, but recently, there is growing interest in analyses for the dynamic data (time-dependent data)~\cite{Puglisi_2013,Puglisi_2014_1,Puglisi_2014_2,Stampfer_2018,Murai_2018,Doi_2016,Tamaoka_2021,Gammelmark_2014,Pirchi_2016,Nguyen-Le_2020}.

In the present paper, we present an analysis for a time-series data on optical properties of single quantum dots (QDs). The QD is a fine particle of several nanometers and is a system that can confine electrons and holes in all dimensions. Due to its unique electronic and fluorescent properties, there are wide applications including single-electron transistor~\cite{Zhuang_1998}, quantum teleportation~\cite{Pirandola_2015}, QD laser~\cite{Huffaker_1998}, QD solar cell~\cite{Nozik_2002}, and quantum computers~\cite{Imamog_1999}. From a viewpoint of the performance of the QD as emiters, it is desirable that the light emission continues as long as possible under continuous excitation, leading to a high performance on luminescence quantum yield. However, it is known that the single QDs show blinking behaviors~\cite{Nirmal_1996}. As a fundamental property for this measure, a distribution function on the light-emission states of the single QD is widely employed and called a blinking plot~\cite{Kuno2001}. 

\begin{figure}[b]
\centering
\includegraphics[width=1.0\linewidth]{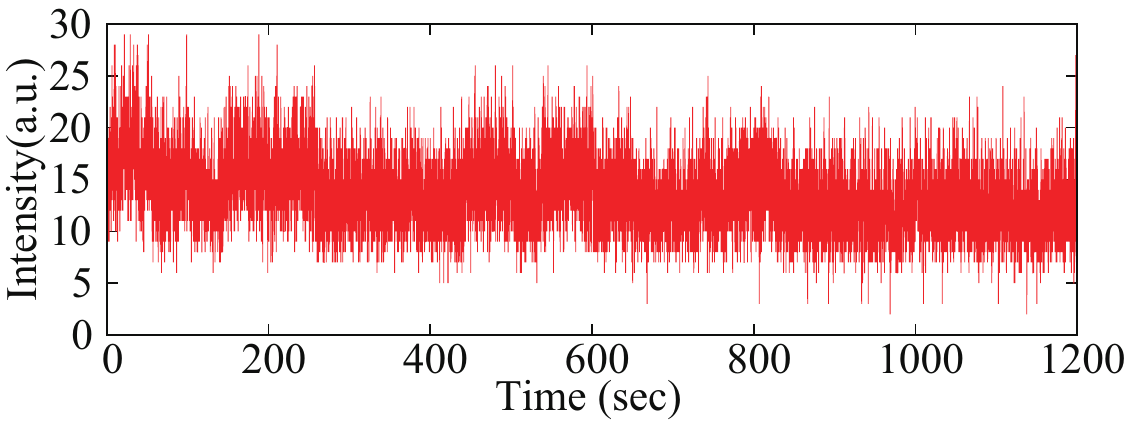}
\caption{Experimental fluorescence time-series data of quantum dots. See Sec.~\ref{Sec_implementation_of_experimental_data} for experimental details and conditions.}
\label{time-series_data}
\end{figure}

In the experiment, we irradiate the QD with the light continuously and observe a situation where the QD emits or quenches in real time. Figure~\ref{time-series_data} shows an example of the observed fluorescent intensities of a single QD as a function of time. As can be seen from the figure, the time series of the fluorescent intensity is noisy due to the experimental equipment and the measurement environment. It is difficult to quantitatively distinguish the light-emission and quenching states especially in such case that the noises are larger or comparable to the fluorescent signals. Conventionally, in order to define the light-emission states, an artificial threshold is introduced by hand in the intensity analysis, and all the states beyond this threshold are evaluated as the light-emission state. With this definition, the light-emission duration are estimated, and a distribution function on the duration is evaluated. With this method, however, the result will include a large amount of short light-emission duration that do not actually exist, and then, the obtained distribution function would clearly be erroneous.

In this study, we present a time-series data analysis using a hidden Markov model (HMM)~\cite{RABINER_1989,ZOUBIN_2001} to solve this problem. The HMM is a typical machine learning method to handle time-series data, and is applied to various dynamic data analyses such as stock price prediction~\cite{HASSAN_2009} and anomaly detection~\cite{KWON_1999}. In material science, it is also beginning to be used for several applications including random telegraph noise~\cite{Puglisi_2013,Puglisi_2014_1,Puglisi_2014_2,Stampfer_2018}, nitrogen-vacancy center in diamond~\cite{Murai_2018,Doi_2016}, electron holograms~\cite{Tamaoka_2021}, atom quantum jump dynamics~\cite{Gammelmark_2014}, fluorescence resonance energy transfer~\cite{Pirchi_2016}, and crack propagation~\cite{Nguyen-Le_2020}. The HMM introduces hidden variables for specifying states behind the observed real time series. We will show that the above mentioned noises in the fluorescence data can be eliminated with the HMM method; the time series of the hidden variable are noise-suppressed, and as a result, the duration evaluation becomes stable. 

The present paper is organized as follows: In Sec.~\ref{sec_method}, we describe a basic idea for treating the present problem, details of the HMM,  and how to analyze the time series data with the HMM. In Sec.~\ref{sec_results_and_discussions}, we present results of the HMM analysis for experimental fluorescence data. On top of that, to verify accuracy of the present HMM analysis, we show performance for the theoretical benchmark data. Section~\ref{sec_summary} summarize conclusions. We also in Appendix~\ref{Blocking Gibbs sampling} describe details for the blocking Gibbs sampling to solve the HMM.  


\section{Method}\label{sec_method} 

\subsection{Basic idea on state identification}\label{Basic_idea_on_State_identification}
\begin{figure}[htb]
\centering
\includegraphics[width=0.8\linewidth]{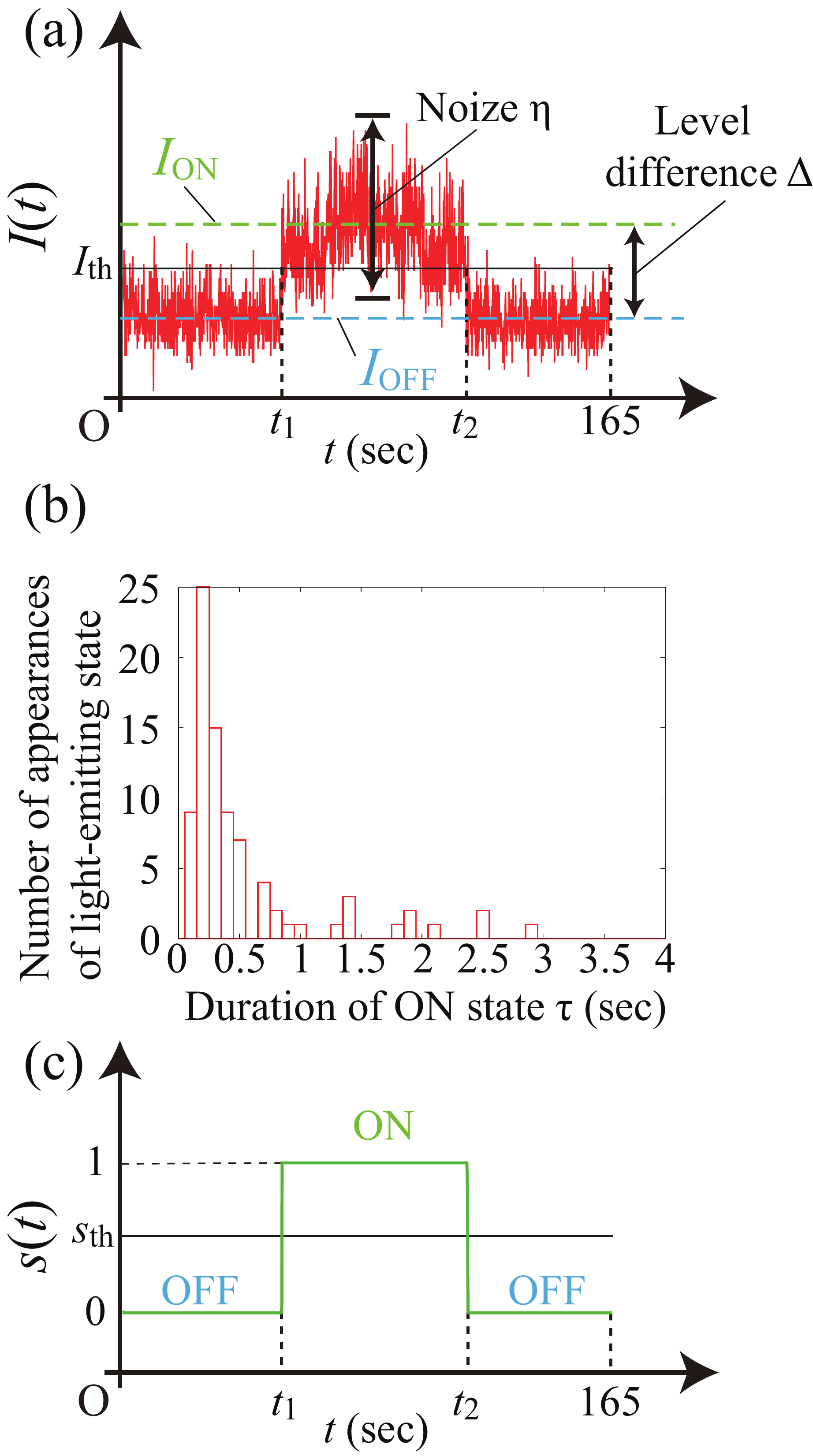}
\caption{Digestive figures on an analysis for fluorescent time series: (a) Time series of the fluorescent intensity $I(t)$, where $I_{{\rm ON}}$ and $I_{{\rm OFF}}$ are the guides representing baselines of intensity for the ON and OFF states. $\Delta$ and $\eta$ are the level difference between the two states and noise, respectively. $I_{th}$ is a threshold to distinguish the ON and OFF states, which is represented by a solid line (see the text). (b) Number of appearances of the ON (bright) state, where we count the appearances by duration for the time series in the panel (a). (c) Time series of a hidden variable $s(t)$ obtained from HMM simulations for the fluorescent time series in the panel (a) (see Sec.~\ref{Hidden Markov model} for details), where $s(t)=1$ and $s(t)=0$ indicate the ON and OFF states, respectively. Note that the noise is totally suppressed in the $s(t)$ time series.}
\label{SchematicFigureStateJudgement}
\end{figure}

Before presenting details, we first show a basic idea for studying a blinking phenomena. As an example, we consider a time series of a fluorescent intensity $I(t)$ in Fig.~\ref{SchematicFigureStateJudgement}~(a). In this time series, the intensity is high at the regions of $t_1 \le t \le t_2$  and low in the range of $0 \le t \le t_1$ and $t_2 \le t$. Here, we define the former region as an ON (bright) state and the latter region as an OFF (dark) state. The time series contains a noise $\eta$, which makes it difficult to determine the ON or OFF state; if the level difference  $\Delta$ between the two states competes with the $\eta$, the state assignment becomes difficult.

In the conventional method, a threshold intensity $I_{th}$ in the panel~(a) is introduced to define the ON and OFF states; if the intensity $I(t)$ is larger than the $I_{th}$, the state at the time $t$ is specified as the ON states. On the other hand, if the intensity $I(t)$ is smaller than the $I_{th}$, the state is classified as the OFF state.  Figure~\ref{SchematicFigureStateJudgement}~(b) shows a statistics on the duration $\tau$ of the ON state, based on the conventional method (see Sec.~\ref{Sec_analysis_for_result_of_HMM} for details). Since the intensity $I(t)$ largely fluctuates due to the noise, many short-duration states of $\tau\sim$ 0.1-0.5 s generate. In the normal sense, however, in the case of this time series, the ON duration would be $t_2-t_1=62.6$ s and there are no shorter ON states. From this discussion, the evaluation of the duration with the conventional method would bring about an erroneous result for the noisy time series.

Next, we consider the state assignment by HMM. The HMM introduces a time series of a hidden variable $s(t)$ that governs the generation of the $I(t)$ (see Sec.~\ref{Hidden Markov model}). When $s(t)$ = 1, the state at the time $t$ is ON associated with the high fluorescent intensity, while when $s(t)$ = 0, the state at the time $t$ is OFF with the low intensity. The $s(t)$ time series can be determined with the Bayesian inference for the observed time series $I(t)$. Note that the noise is suppressed in the $s(t)$ time series [Fig.~\ref{SchematicFigureStateJudgement}~(c)]. Since the $s(t)$ has basically a binary feature, by introducing a threshold $s_{th}$ in the panel~(c), we can perform a stable state assignment.

\subsection{Hidden Markov model}\label{Hidden Markov model}
The HMM is in general expressed with a combination of various probability distributions of stochastic variables~\cite{RABINER_1989,ZOUBIN_2001}. A form of the HMM employed in the present study is shown below: 
\begin{eqnarray}
 p({\bm I}, \mathbf{S}, \mathbf{\Theta}, \bm{\pi}, \mathbf{A}) 
 =p({\bm I}|\mathbf{S}, \mathbf{\Theta}) 
 p(\mathbf{S}|\bm{\pi},\mathbf{A}) 
 p(\mathbf{\Theta}) 
 p(\bm{\pi}) 
 p(\mathbf{A}). \nonumber \\
 \label{douji-2}
\end{eqnarray}
Here, ${\bm I}$ is observed data, $\mathbf{S}$ is hidden variable, $\mathbf{\Theta}$ describes parameters characterizing the distribution function of ${\bm I}$, ${\bm \pi}$ is parameters to set the initial step of the hidden variable, and $\mathbf{A}$ describes a transition matrix for the time evolution of hidden variables. The $p(\mathbf{S}|\bm{\pi}, \mathbf{A})$ in Eq.~(\ref{douji-2}) describes the conditional probability distribution of $\mathbf{S}$ with $\mathbf{A}$ and $\bm{\pi}$ fixed, and similarly, $p(\bm{I}|\mathbf{S},\mathbf{\Theta})$ is the conditional probability distribution of $\bm{I}$ after $\mathbf{S}$ and $\mathbf{\Theta}$ are determined. 

${\bm I}$ in Eq.~(\ref{douji-2}) is discrete time-series data written as
\begin{eqnarray}
 {\bm I}=(I_1, I_2, \cdots, I_N)^T, 
 \label{intensity} 
\end{eqnarray}
where $I_n$ is an observed value at the $n$-th time grid, and the total number of grids is $N$. In the present study, the ${\bm I}$ corresponds to the time series of the fluorescent intensity of a single QD.
  
$\mathbf{S}$ in Eq.~(\ref{douji-2}) is hidden variables written as 
  \begin{eqnarray}
   \mathbf{S}=(\mathbf{s}_1, \mathbf{s}_2, \cdots, \mathbf{s}_N), 
   \label{time-series-data-for-s}
  \end{eqnarray}
which is also a time series. The component ${\bf s}_n$ is a vector and consists of $K$ components as
  \begin{eqnarray}
   \mathbf{s}_n=(s_{1n},s_{2n},\cdots,s_{Kn})^T.
   \label{s_components}
  \end{eqnarray}
The ${\bf s}_n$ describes an internal state behind the fluorescence data at the time $t_n$. In the present study, the number of the internal states is 2 ($K=2$), and thus, the ${\bf s}_n$ is described with binary as   
\begin{eqnarray}
 \mathbf{s}_{n}=
 \begin{cases}
  \begin{pmatrix}
    1 \\
    0   
  \end{pmatrix} & ({\rm ON\ state}), \vspace{0.3cm} \\  
  \begin{pmatrix}
    0 \\
    1   
  \end{pmatrix} & ({\rm OFF\ state}).  
 \end{cases}
 \label{s_kn}
\end{eqnarray}
Note that this variable satisfies a sum rule as 
\begin{eqnarray}
 \sum_{k=1}^{K}s_{kn}=1. 
 \label{sumrule-s}
\end{eqnarray}

$\bm{\Theta}$ in Eq.~(\ref{douji-2}) expresses variables to characterize a probability distribution function of the time series ${\bm I}$ in Eq.~(\ref{intensity}). Since the total number of the internal states is $K$, we consider $K$ distribution functions and prepare parameters for each distribution function. In the present study, we assume that the distribution function of each state follows the Gaussian distribution. In the Gaussian function, it is characterized by mean $\mu$ and precision $\lambda$, where $\lambda^{-1}$ represents a variance. Thus, we rewrite ${\bm \Theta}$ as $\{{\bm \mu}, {\bm \lambda}\}$, where $\bm{\mu}$ contains the mean of each state as
\begin{eqnarray}
 \bm{\mu}=(\mu_1, \mu_2, \cdots, \mu_K)^T 
 \label{mu_vec}
\end{eqnarray}
and $\bm{\lambda}$ contains the precision of each state as  
\begin{eqnarray}
 \bm{\lambda}=(\lambda_1, \lambda_2, \cdots, \lambda_K)^T. 
 \label{lambda_vec}
\end{eqnarray}
Here, $\mu_k$ and $\lambda_k$ are the mean and precision of the $k$-th Gaussian distribution.

$\bm{\pi}$ in Eq.~(\ref{douji-2}) describes parameters to set an initial-step hidden variable $\mathbf{s}_1$, and this is also a $K$-component vector as
\begin{eqnarray}
 \bm{\pi}=(\pi _1, \pi _2, \cdots, \pi _K)^T 
 \label{pi_vec}
\end{eqnarray}
with $\pi _k$ being the probability of taking the $k$-th state at the initial step. Note a normalization of $\sum_{k=1}^K \pi_k =1$. 
 
$\mathbf{A}$ in Eq.~(\ref{douji-2}) is variables for describing the transition between successive hidden variables, which are expressed as  
\begin{eqnarray}
 \mathbf{A}=(\mathbf{A}_1, \mathbf{A}_2, \cdots, \mathbf{A}_K)
 \label{A_mat}
\end{eqnarray}
with 
 \begin{eqnarray}
  \mathbf{A}_{k'}=(A_{1k'}, A_{2k'}, \cdots, A_{Kk'})^T
 \end{eqnarray}
representing a transition probability from $k'$-th state to other states.  Thus, $\mathbf{A}$ is a $K \times K$ matrix, and the matrix element $A_{kk'}$ expresses a transition probability from the $k'$-th state to the $k$-the state. Note the normalization of $\sum_{k=1}^K A_{kk'}=1$.

\begin{figure}[htb]
\centering
\includegraphics[width=\linewidth]{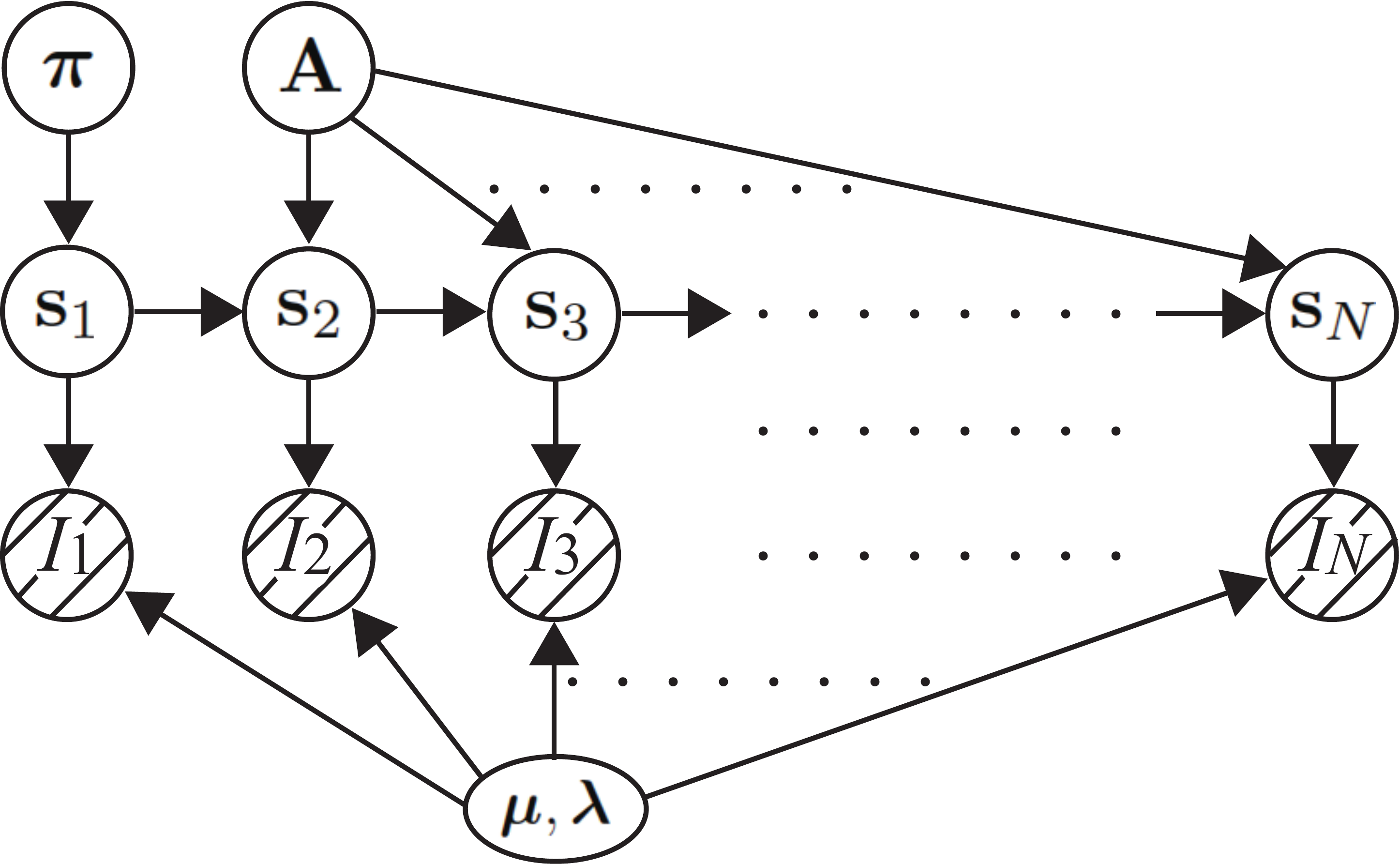}
\caption{A graphical model for HMM defined as Eq.~(\ref{douji-3}): \{$I_n$\} are observed time series [Eq.~(\ref{intensity})] and the observed data is drawn with a shadow. \{$\mathbf{s}_n$\} is a time series of hidden variables [Eq.~(\ref{time-series-data-for-s})]. $\bm{\mu}$ and $\bm{\lambda}$ in Eqs.~(\ref{mu_vec}) and (\ref{lambda_vec}) are parameters characterizing distribution functions that time series ${\bm I}$ follows. An initial-step hidden variable $\mathbf{s}_1$ is generated based on the parameter $\bm{\pi}$ in Eq.~(\ref{pi_vec}), while $\mathbf{s}_n$ is generated by the previous-step $\mathbf{s}_{n-1}$ and the transition matrix $\mathbf{A}$ in Eq.~(\ref{A_mat}). Stochastic variables are represented with nodes and the dependency between the variables is represented by arrows.}
\label{graphical_model}
\end{figure}

We now rewrite the right-hand side of Eq.~(\ref{douji-2}) by considering independency of the stochastic variables as~\cite{RABINER_1989,ZOUBIN_2001}   
\begin{eqnarray}
 & &p({\bm I}|\mathbf{S}, \bm{\mu}, \bm{\lambda}) 
 p(\mathbf{S}|\bm{\pi},\mathbf{A}) 
 p(\bm{\mu},\bm{\lambda}) 
 p(\bm{\pi}) 
 p(\mathbf{A}) \nonumber \\ 
 & &= 
 \Biggl(\prod_{n=1}^N \prod_{k=1}^K p(I_n|\mu_k,\lambda_k)^{s_{kn}}\Biggr) \nonumber \\
 & &\times \Biggl(p(\mathbf{s}_1|\bm{\pi})\prod_{n=2}^N\prod_{k'=1}^K  
 p(\mathbf{s}_{n}|\mathbf{A}_{k'})^{s_{k'n-1}}\Biggr) \nonumber \\
 & &\times \Biggl(\prod_{k=1}^K p(\mu_{k},\lambda_{k})\Biggr)
 p(\bm{\pi})
 \Biggl(\prod _{k'=1}^K p(\mathbf{A}_{k'})\Biggr). 
\label{douji-3}
\end{eqnarray}
Here, we note that the forms of $\prod_{k=1}^K p(I_n|\mu_k,\lambda_k)^{s_{kn}}$ and $\prod_{k'=1}^K p(\mathbf{s}_{n}|\mathbf{A}_{k'})^{s_{k'n-1}}$ in Eq.~(\ref{douji-3}) select only one of the $K$ 
states due to the binary feature of the hidden variable $\mathbf{s}_n$ [Eqs.~(\ref{s_kn}) and (\ref{sumrule-s})]. The variable dependency in this model is represented as a graph in Fig.~\ref{graphical_model}. Arrows represent a dependency between variables; for example, in this model, $I_n$ depends on $\mathbf{s}_n$, $\bm{\mu}$, and $\bm{\lambda}$, which corresponds to $\prod_{k=1}^K p(I_n|\mu_k,\lambda_k)^{s_{kn}}$ of the right-hand side in Eq.~(\ref{douji-3}). 

Concrete forms of each probability distribution in Eq.~(\ref{douji-3}) is summarized as follows: 
\begin{enumerate}
 \item 
 $p(I_n|\mu_k,\lambda_k)$ is the Gaussian distribution as  
 \begin{eqnarray}
 p(I_n|\mu_k,\lambda_k)&=&\mathcal{N}(I_n|\mu_k,\lambda_k^{-1}) \nonumber \\
  &=&\sqrt{\frac{\lambda_k}{2 \pi}} \exp{\biggl(-\frac{\lambda_k}{2}(I_n-\mu_k)^2}\biggr). \label{Gauss}
 \end{eqnarray}
 
 \item $p(\mathbf{s}_1|\bm{\pi})$ and $p(\mathbf{s}_{n}|\mathbf{A}_{k'})$ are the categorical distribution as  
 \begin{eqnarray}
 p(\mathbf{s}_1|\bm{\pi})&=&\mathrm{Cat}(\mathbf{s}_1|\bm{\pi}) 
 = \prod_{k=1}^{K} \pi_{k}^{s_{k1}}
 \end{eqnarray}
 and 
 \begin{eqnarray}
 p(\mathbf{s}_{n}|\mathbf{A}_{k'})&=&\mathrm{Cat}(\mathbf{s}_{n}|\mathbf{A}_{k'}) 
 = \prod_{k=1}^{K} A_{kk'}^{s_{kn}},
 \end{eqnarray}
 respectively.  
 
 \item $p(\mu_{k},\lambda_{k})$ is the Gaussian-Gamma distribution as  
 \begin{eqnarray}
  p(\mu_{k},\lambda_{k})
  &=&{\rm NG}(\mu_{k},\lambda_{k}|m,\nu,a,b) \notag \\
  &=&\mathcal{N}(\mu_{k}|m,(\nu \lambda_{k})^{-1}) \mathrm{Gam}(\lambda_{k}|a,b) \notag \\
  &=& \Biggl \{\sqrt{\frac{\nu\lambda_{k}}{2\pi}} \exp{\biggl(-\frac{\nu\lambda_{k}}{2}(\mu_{k}-m)^2}\biggr)\Biggr\} \nonumber \\
  &\times& \Biggl \{C_G(a,b)\lambda_{k}^{a-1}e^{-b\lambda_{k}} \Biggr \},
  \label{gauss_ganmma_distribution}
 \end{eqnarray}
 where $m$, $a$, $b$, and $\nu$ are hyperparameters in the Gaussian-Gamma distribution, and $C_G(a,b)$ is the normalization constant as 
 \begin{eqnarray}
  C_G(a,b)=\frac{b^a}{\Gamma (a)}
 \end{eqnarray} 
 with $\Gamma(a)$ being the Gamma function. 

 \item $p(\bm{\pi})$ and $p(\mathbf{A}_{k'})$ are the Dirichlet distribution as 
 \begin{eqnarray}
  p(\bm{\pi})&=&\mathrm{Dir}(\bm{\pi}|\bm{\alpha})
  = C_D(\bm{\alpha})\prod _{k=1}^K \pi_k^{\alpha_k-1}  \label{pi_Dirichlet_distribution}
 \end{eqnarray}
 and  
 \begin{eqnarray}
  p(\mathbf{A}_{k'})&=&\mathrm{Dir}(\mathbf{A}_{k'}|\bm{\beta}_{k'})=C_D(\bm{\beta}_{k'})\prod _{k=1}^K A_{kk'}^{\beta_{kk'}-1}, 
  \label{A_Dirichlet_distribution}
 \end{eqnarray}
 respectively, and $\bm{\alpha}$ and $\bm{\beta}$ are hyperparameters in the Dirichlet distribution. Also, $C_D(\bf{x})$ is the normalization constant written as 
 \begin{eqnarray}
  C_D(\mathbf{x})=\frac{\Gamma (\sum_{k=1}^{K} x_k)}{\prod_{k=1}^{K} \Gamma (x_k)}.
  \end{eqnarray}
\end{enumerate}

In the practical simulation, we need to calculate the conditional distribution function under the observed time-series data $\bm{I}$ in Eq.~(\ref{intensity}) as   
 \begin{eqnarray}
  p(\mathbf{S},{\bm \mu}, {\bm \lambda}, \bm{\pi},  \mathbf{A}|\bm{I})
  =\frac{p(\bm{I}, \mathbf{S}, \bm{\mu}, \bm{\lambda}, \bm{\pi}, \mathbf{A})}{p(\bm{I})}, 
  \label{conditional-p}
  \end{eqnarray}
where $p(\bm{I})$ is a marginal distribution written as
 \begin{eqnarray}
  p(\bm{I})
  =\sum_{\mathbf{S}} \iiiint p(\bm{I}, \mathbf{S}, \bm{\mu}, \bm{\lambda}, \bm{\pi}, \mathbf{A}) d\bm{\mu} d\bm{\lambda} d\bm{\pi} d\mathbf{A}. \nonumber \\ 
  \label{pI}
 \end{eqnarray}
In the present study, we used the blocking Gibbs sampling~\cite{JENSEN_1995} based on the Bayesian inference to optimize the conditional distribution function in Eq.~(\ref{conditional-p}). Details can be found in Appendix~\ref{Blocking Gibbs sampling}. Anyway, by evaluating the conditional distribution function, the time-series data of the hidden variable $\mathbf{S}$ in Eq.~(\ref{time-series-data-for-s}) can be obtained, which is used for the following analysis of the ON/OFF duration. 

\subsection{Duration estimate based on HMM} \label{Sec_analysis_for_result_of_HMM}
We next describe an evaluation of the duration of the ON or OFF state of QDs. We first consider a sampling average for the time-series data of the hidden variable as 
\begin{eqnarray}
 \bar{\mathbf{S}}=\frac{1}{N_{itr}} \sum _{i=1} ^{N_{itr}} \mathbf{S}^{(i)},  
 \label{s_i_max}
\end{eqnarray}
where $\mathbf{S}^{(i)}$ is the time-series data of the hidden variable in the $i$-th Gibbs sampling step, and $N_{itr}$ is the total number of the Gibbs sampling. The $\bar{\mathbf{S}}$ is given in the same form as in Eq.~(\ref{time-series-data-for-s}): 
\begin{eqnarray}
 \bar{\mathbf{S}}=(\bar{\mathbf{s}}_1, \bar{\mathbf{s}}_2, \cdots, \bar{\mathbf{s}}_N).
 \label{Save}
\end{eqnarray}
In the present study, the $\mathbf{s}_n$ has two components; the ON and OFF states. Therefore, the $\bar{\mathbf{s}}_n$ is written as  
\begin{eqnarray}
 \bar{\mathbf{s}}_n = 
 \begin{pmatrix}
    s^{{\rm ON}}_n \vspace{0.2cm} \\
    s^{{\rm OFF}}_n   
  \end{pmatrix}
  =
   \begin{pmatrix}
    s^{{\rm ON}}_n \vspace{0.2cm} \\
    1-s^{{\rm ON}}_n   
  \end{pmatrix} 
 \label{s_ON} 
\end{eqnarray}
with $s_n^{{\rm ON}}=(1/N_{itr})\sum_{i}^{N_{itr}} s_{1n}^{(i)}$ and $s_n^{{\rm OFF}}=(1/N_{itr})\sum_{i}^{N_{itr}} s_{2n}^{(i)}$. Also, the sum rule in Eq.~(\ref{sumrule-s}) was used in the transformation from the middle to right-side equations. We note that the component $s^{{\rm ON}}_n$ is a real number, not a binary, after the ensemble average. Now, using $s^{{\rm ON}}_n$, we define a time series $\mathbf{s}^{{\rm ON}}$ as 
\begin{eqnarray}
 \mathbf{s}^{{\rm ON}}=(s^{{\rm ON}}_1, s^{{\rm ON}}_2, \cdots, s^{{\rm ON}}_N)^T.
 \label{sON_vec}
\end{eqnarray}

\begin{figure}[htb]
\centering
\includegraphics[width=\linewidth]{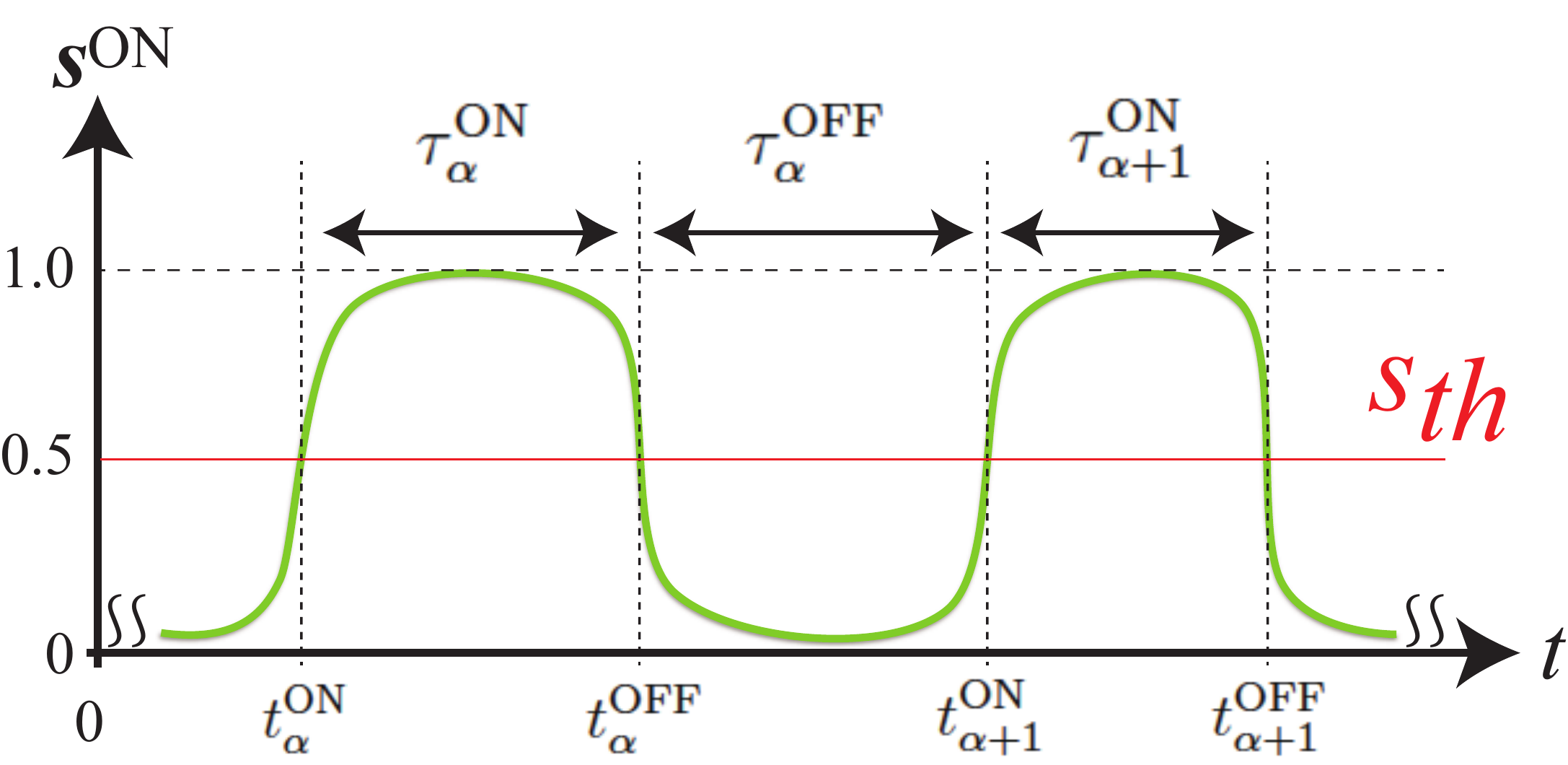}
\caption{A schematic diagram of a time series of $\mathbf{s}^{{\rm ON}}$ [Eq.~(\ref{sON_vec})] which is described by a thick green-solid curve. $s_{th}$ is a threshold to distinguish the ON and OFF states, which is represented by a thin solid-red line. Since the ${\bf s}^{{\rm ON}}$ takes a value from 0 to 1, $s_{th}=0.5$ is adopted. It is defined as entering the ON state when the ${\bf s}^{{\rm ON}}$ curve exceeds the value of $s_{th}$. We set this time to $t^{{\rm ON}}_{\alpha}$ with a suffix $\alpha$ specifying the number of the ON event. Similarly, when the ${\bf s}^{{\rm ON}}$ curve becomes smaller than $s_{th}$, it is defined as entering the OFF state. We set this time to $t^{{\rm OFF}}_{\alpha}$. From these two times, the ON and OFF duration, $\tau^{{\rm ON}}_{\alpha}$ and $\tau^{{\rm OFF}}_{\alpha}$, are estimated with using Eqs.~(\ref{tauon}) and (\ref{tauoff}), respectively.}
\label{duration_time}
\end{figure}

Figure~\ref{duration_time} shows a schematic diagram of $\mathbf{s}^{{\rm ON}}$ denoted by a thick green-solid curve. Generally, in the $\mathbf{s}^{{\rm ON}}$, statistical noise is suppressed and smooth behavior is obtained (see Sec.~\ref{sec_results_and_discussions}). We thus estimate the ON duration $\tau^{{\rm ON}}$ and the OFF duration $\tau^{{\rm OFF}}$ from the $\mathbf{s}^{{\rm ON}}$ data, where a threshold $s_{th}$ denoted by thin red-solid line is introduced. Since the ${\bf s}^{{\rm ON}}$ takes a value from 0 to 1, $s_{th}=0.5$ is adopted. It is defined as entering the ON state when the ${\bf s}^{{\rm ON}}$ curve exceeds the $s_{th}$. We set this time to $t^{{\rm ON}}_{\alpha}$ with a suffix $\alpha$ to specify the number of the ON event. On the other hand, when the ${\bf s}^{{\rm ON}}$ curve becomes smaller than the $s_{th}$, it is defined as entering the OFF state. We set this time to $t^{{\rm OFF}}_{\alpha}$. From these two times, the duration of the ON state is defined as 
\begin{eqnarray}
 \tau^{{\rm ON}}_{\alpha} = t^{{\rm OFF}}_{\alpha} - t^{{\rm ON}}_{\alpha}. 
 \label{tauon}
\end{eqnarray}
Similarly, we define the duration of the OFF state as
\begin{eqnarray}
 \tau^{{\rm OFF}}_{\alpha} = t^{{\rm ON}}_{\alpha+1} - t^{{\rm OFF}}_{\alpha}.
 \label{tauoff} 
\end{eqnarray}

With the obtained \{$\tau^{{\rm ON}}_{\alpha}$\} and \{$\tau^{{\rm OFF}}_{\alpha}$\} data, probability distributions of the ON and OFF duration are calculated by 

\begin{eqnarray}
 P_{{\rm ON}}(\tau)
 =\frac{1}{N_{{\rm ON}}}
  \sum_{\alpha'=1}^{N^{\prime}_{{\rm ON}}}
       f_{\alpha'}^{{\rm ON}} w_{\alpha'}^{{\rm ON}} \delta(\tau-\tau_{\alpha'}^{{\rm ON}}) 
       \label{dens_ON} 
\end{eqnarray}
and 
\begin{eqnarray}
 P_{{\rm OFF}}(\tau)
 =\frac{1}{N_{{\rm OFF}}}
 \sum_{\alpha'=1}^{N^{\prime}_{{\rm OFF}}}
       f_{\alpha'}^{{\rm OFF}} w_{\alpha'}^{{\rm OFF}} \delta(\tau-\tau_{\alpha'}^{{\rm OFF}}), 
       \label{dens_OFF}
\end{eqnarray}
respectively, where $N^{\prime}_{{\rm ON}}$ and $N^{\prime}_{{\rm OFF}}$ are total numbers of independent ON and OFF events, respectively. We note that, in this calculation, the two duration are considered equal within the grid spacing $\Delta\tau$. $f_{\alpha'}^{{\rm ON}}$ and $f_{\alpha'}^{{\rm OFF}}$ are the number of events with the same ON and OFF duration, respectively. Also, $w_{\alpha'}^{{\rm ON}}$ and $w_{\alpha'}^{{\rm OFF}}$ are weights given as~\cite{Kuno2001}
\begin{eqnarray}
w_{\alpha'}^{{\rm ON}}
=\frac{2 \Delta \tau}{\tau_{\alpha'+1}^{{\rm ON}}-\tau_{\alpha'-1}^{{\rm ON}}} 
\end{eqnarray}
and
\begin{eqnarray}
w_{\alpha'}^{{\rm OFF}}=\frac{2 \Delta \tau}{\tau_{\alpha'+1}^{{\rm OFF}}-\tau_{\alpha'-1}^{{\rm OFF}}},
\end{eqnarray}
respectively. Lastly, the denominators $N_{{\rm ON}}$ and $N_{{\rm OFF}}$ in Eq.~(\ref{dens_ON}) and Eq.~(\ref{dens_OFF}) are the total number of the ON and OFF events, which satisfy the following sum rules:
\begin{eqnarray}
 N_{{\rm ON}}
 =\sum_{\alpha'=1}^{N^{\prime}_{{\rm ON}}} f_{\alpha'}^{{\rm ON}}\frac{w_{\alpha'}^{{\rm ON}}}{S^{{\rm ON}}},
\end{eqnarray}
and
\begin{eqnarray}
 N_{{\rm OFF}}
 =\sum_{\alpha'=1}^{N^{\prime}_{{\rm OFF}}} f_{\alpha'}^{{\rm OFF}}\frac{w_{\alpha'}^{{\rm OFF}}}{S^{{\rm OFF}}},
\end{eqnarray}
respectively, where we introduced correction factors $S^{{\rm ON}}$ and $S^{{\rm OFF}}$ which are determined from the normalization condition of  $P_{{\rm ON}}(\tau)$ and $P_{{\rm OFF}}(\tau)$ .   


\section{Results and Discussions}\label{sec_results_and_discussions} 
We implemented the method described in Sec.~\ref{sec_method} into the program code written by {\sc Python}. With this program, we performed HMM simulations for experimental and theoretical benchmark data. TABLE~\ref{table-HMM} summarizes the present setting of our HMM simulation and the employed hyperparameters. In the simulations, observed fluorescence data $\bm{I}$ in Eq.~(\ref{intensity}) is standardized as 
\begin{eqnarray}
      \tilde{I}_n = \frac{I_n - \bar{I}}{\sigma_I}
\end{eqnarray}
with a mean of  
\begin{eqnarray}
    \bar{I}=\frac{1}{N} \sum_{n=1}^{N} I_n \label{I_mean}
\end{eqnarray}
and a standard deviation of  
\begin{eqnarray}
    \sigma_I = \sqrt{\frac{1}{N}\sum_{n=1}^{N}(I_n-\bar{I})^2}.
\end{eqnarray}
The standardized time series $\tilde{\bm{I}}$ has a mean value of 0 and a standard deviation of 1, which is used as input as the HMM simulations.
\begin{table}[!]
\caption{Initial hyperparameter setting for the present HMM simulations. $K$ is the number of the hidden states, and $\nu$, $m$, $a$, and $b$ are the hyperparameters for the Gaussian-Gamma distribution of $p(\mu_{k},\lambda_{k})$ in Eq.~(\ref{gauss_ganmma_distribution}). $\bm{\alpha}$ is the hyperparameter for the distribution function $p(\bm{\pi})$ in Eq.~(\ref{pi_Dirichlet_distribution}), which is a $K$-component vector. Also, $\bm{\beta}$ is the hyperparameter for the distribution function $p(\mathbf{A}_{k'})$ in Eq.~(\ref{A_Dirichlet_distribution}), which is represented by a $K\times K$ matrix. The matrix elements are characterized by $N$, $D$, and $\gamma$, where $N$ is the total length of the time series, and $D$ and $\gamma$ were set to 1 and 0.01 in the present simulation, respectively. 
$N_{itr}$ is the total number of the iteration steps of the Gibbs sampling, which depends on the simulations.} 
\begin{center}
\scalebox{0.93}{
\begin{tabular}{lr} 
\hline \hline \\ [-5pt] 
$K$ in Eq.~(\ref{s_components}) & 2\\ [5pt] 
$(\nu,m, a, b)$ in  Eq.~(\ref{gauss_ganmma_distribution}) & $(1, 0, 1, 1)$ \\ [5pt] 
$\bm{\alpha}$ in Eq.~(\ref{pi_Dirichlet_distribution})   & $(100,100)^T$ \\ [10pt] 
$\bm{\beta}$ in Eq.~(\ref{A_Dirichlet_distribution})    &  
$\Biggl( \begin{array}{cc} \frac{(1-2\gamma)N+4(1-\gamma)D}{4\gamma} & D \\ D & \frac{(1-2 \gamma )N + 4(1-\gamma)D}{4 \gamma} \end{array} \Biggr)$  \\ [20pt] 
$N_{itr}$ in Eq.~(\ref{s_i_max}) & 
$\left\{ 
 \begin{array}{cl} 
  1000 & \mbox{[Expt. (Figs.~\ref{implementation_of_experimental_data_in_vac}, \ref{implementation_of_experimental_data_in_wet})}] \\ [3pt] 
  200  & \mbox{[TS-I and TS-II (Figs.~\ref{photon_signal_time-series_data}~(a), (b))]} \\ [3pt]
  1000 & \mbox{[TS-III (Fig.~\ref{photon_signal_time-series_data}~(c))]} 
 \end{array} 
 \right.$  \\ [15pt] 
 \hline \hline
\end{tabular} 
}
\end{center}
\label{table-HMM}
\end{table}

\subsection{HMM analysis for experimental data}\label{Sec_implementation_of_experimental_data}

\subsubsection{Experimental fluorescence data}\label{sec_expt_condition} 
\begin{figure}[!]
\centering
\includegraphics[width=1.0\linewidth]{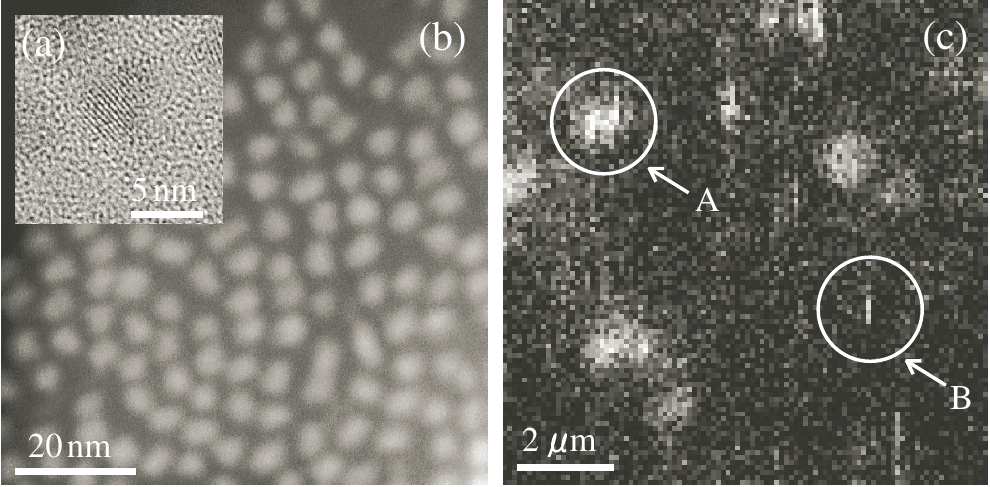}
\caption{(a) Transmission electron microscope (TEM) and (b) scanning TEM images. These images were taken with a FEI Tecnai 20 TEM. (c) Scanning fluorescence image of the QDs. The circled areas A and B in the panel show examples of ON (bright) and OFF (dark) states in the QDs, respectively.}
\label{TEM_SEM_scanning_images}
\end{figure}

Here, we apply the above method to analysis of the experimental data. Before presenting the analysis results, we first describe experimental detalils. The QDs used in the experiment are CdSe/ZnS core-shell QDs capped with trioctylphosphine oxide and hexadecyl amine, synthesized via the pyrolytic decomposition of organometallic compounds~\cite{HASHIZUME_2002}. The QDs have a spheroid shape (minor axis: 4.1 $\pm$ 1.2 nm; major axis: 5.3 $\pm$ 1.3 nm), as confirmed in the transmission electron microscope (TEM) [Fig.~\ref{TEM_SEM_scanning_images}~(a)] and the scanning TEM [Fig.~\ref{TEM_SEM_scanning_images}~(b)] images. The thickness of the ZnS shell is 1.7 monolayers. The QDs were spin-cast from a toluene solution at 3,000 rpm onto silica glass plates with a hydrophobic surface. The spin-cast samples were set in a sealed chamber with optical windows.

We measured scanning fluorescence images and time series of a fluorescent intensity of single QDs using a scanning laser microscope~\cite{ODA_2007} equipped with an objective lens (100$\times$; 0.9 NA) and avalanche photodetector. The laser-light wavelength and excitation intensity were 488 nm and 133 W/cm$^2$, respectively. The fluorescence images were acquired by raster scanning a laser spot on the sample surface. The laser spot size and scanning step length were the diffraction limited size (less than 1 $\mu$m) and 0.1 $\mu$m, respectively. An exposure time for each step was 0.1 s. The time series data of the single QDs were recorded at fixed positions on the sample surface with an exposure time of 0.1 s under continuous laser illumination. The fluorescence images and the time series were measured for two different environments; vacuum and wet-nitrogen atmospheres. The wet-nitrogen atmosphere was prepared by bubbling nitrogen gas through distilled water. We took 12 samples with a length of 1200 s for each of the vacuum and wet-nitrogen atmospheres. 

Figure~\ref{TEM_SEM_scanning_images} (c) shows a typical fluorescence image of the spin-cast sample in vacuum. There are several circular spots like A with diffraction limited size and several line segments like B along the scanning direction (vertical direction). All the circular spots and the line segments correspond to the fluorescence images of individual single QDs; the circular spots indicate that the QDs were in the ON state during the raster scanning, while the line segments indicate that the QDs were mainly in the OFF state and temporally in the ON state. All time series data of single QDs shown below were recorded at the center positions of the bright spots in scanning fluorescence images.

\begin{figure*}[!]
\centering
\includegraphics[width=\linewidth]{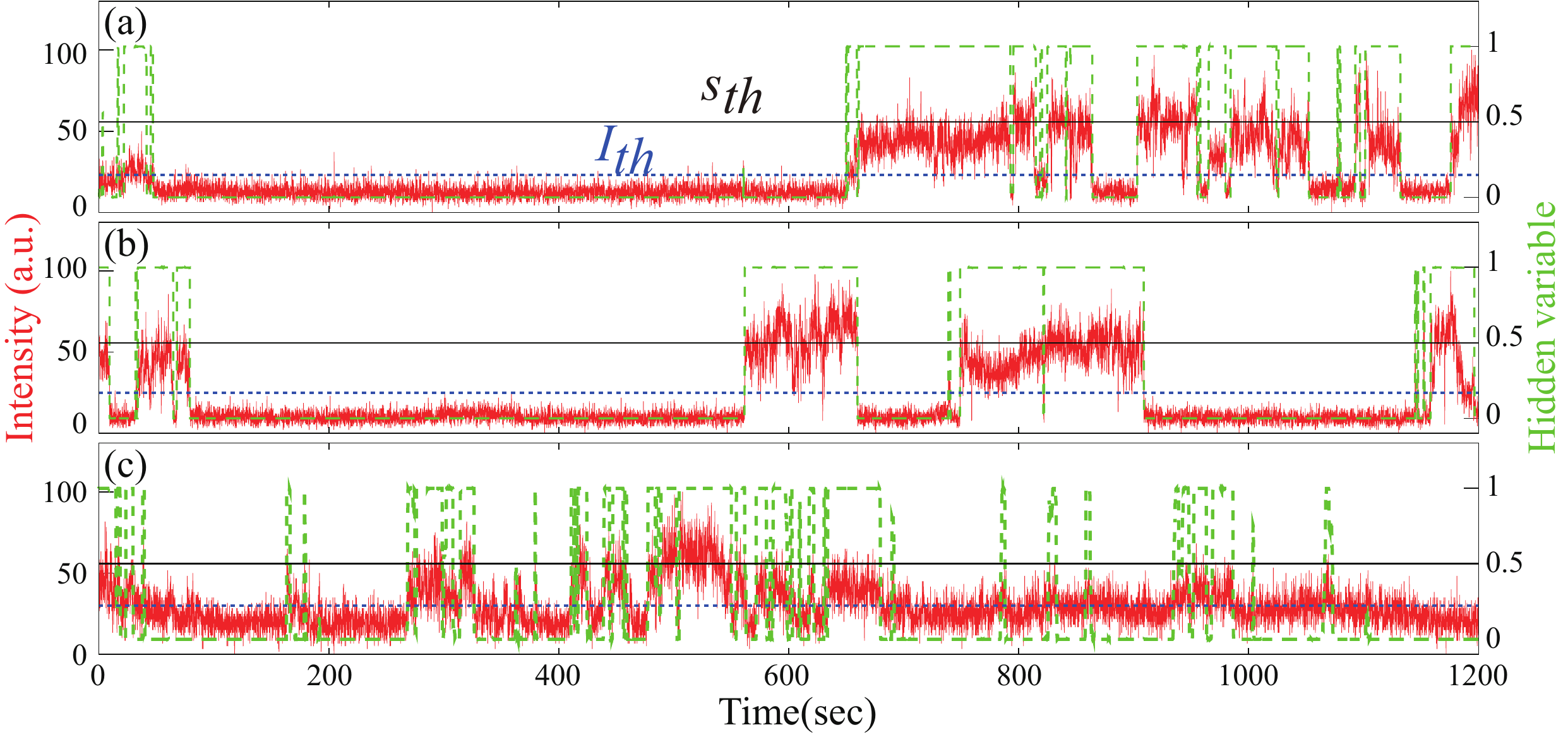}
\caption{Typical three time series of the fluorescent intensity measured in a vacuum atmosphere, denoted by red-solid lines. These are used in the HMM simulations as the observed time series ${\bm I}$ in Eq.~(\ref{intensity}). The grid spacing of the time series is 0.1 s. The green-dashed lines describe the time series of the hidden variable $\mathbf{s}^{{\rm ON}}$ in Eq.~(\ref{sON_vec}) obtained from the HMM simulations. The left scale is for the fluorescent intensity and the right scale is for the hidden variable. The $I_{th}$ denoted by blue-dotted lines is a threshold to distinguish the ON (bright) and OFF (dark) states for the ${\bm I}$ time series, and is set by hand artificially. The $s_{th}$ denoted by black-solid line is a threshold for the $\mathbf{s}^{{\rm ON}}$ time series, and is set to 0.5.} 
\label{implementation_of_experimental_data_in_vac}
\end{figure*}

\begin{figure*}[!]
\centering
\includegraphics[width=\linewidth]{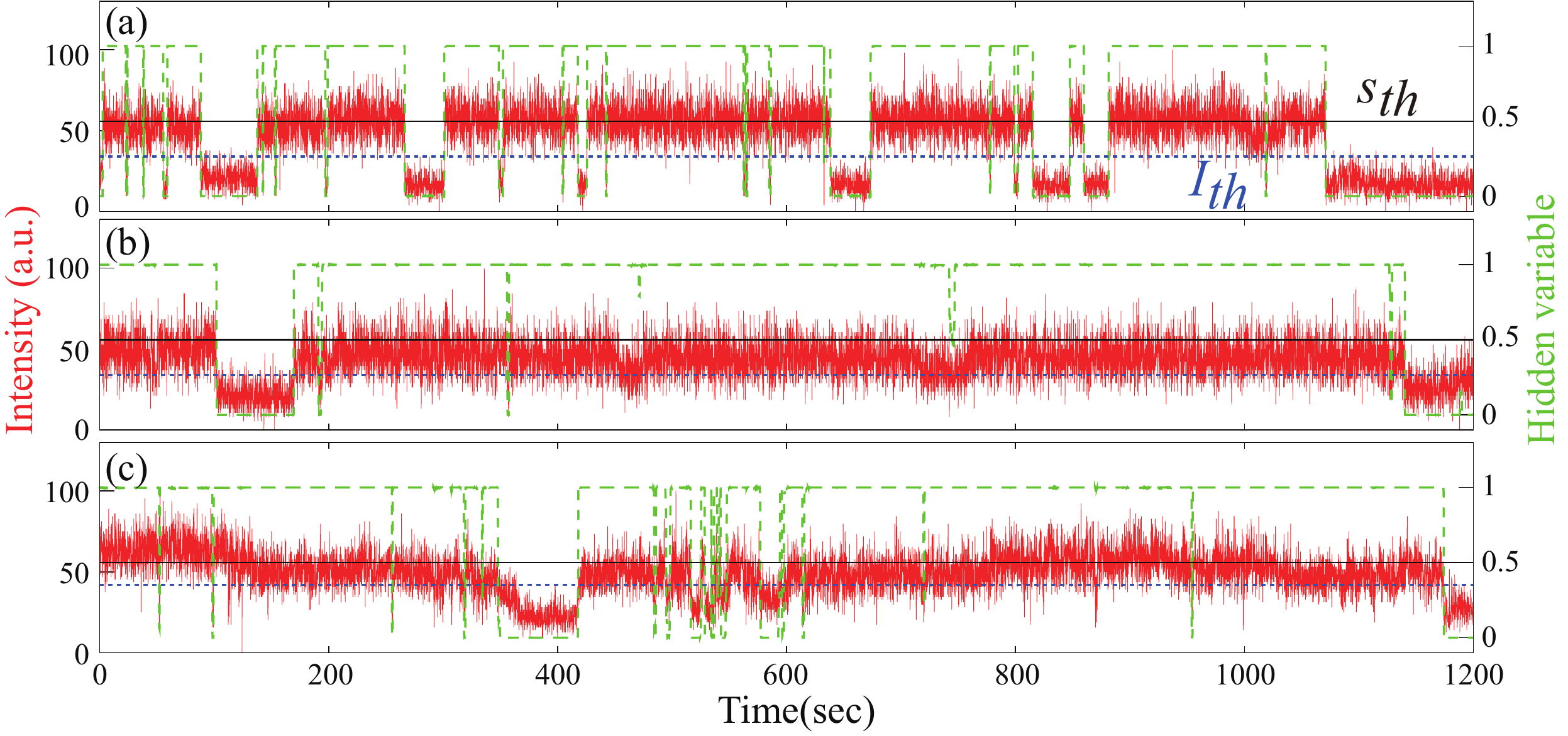}
\caption{Typical three time series of the fluorescent intensity measured in a wet-nitrogen atmosphere, denoted by red-solid lines. These are used in the HMM simulations as the observed time series ${\bm I}$ in Eq.~(\ref{intensity}). The grid spacing of the time series is 0.1 s. The green-dashed lines describe the time series of the hidden variable $\mathbf{s}^{{\rm ON}}$ in Eq.~(\ref{sON_vec}) obtained from the HMM simulations. The left scale is for the fluorescent intensity and the right scale is for the hidden variable. The $I_{th}$ denoted by blue-dotted lines is a threshold to distinguish the ON (bright) and OFF (dark) states for the ${\bm I}$ time series, and is set by hand artificially. The $s_{th}$ denoted by black-solid line is a threshold for the $\mathbf{s}^{{\rm ON}}$ time series, and is set to 0.5.}
\label{implementation_of_experimental_data_in_wet}
\end{figure*}

Figures~\ref{implementation_of_experimental_data_in_vac} (a)-(c) and \ref{implementation_of_experimental_data_in_wet} (a)-(c) show typical time series of the experimental fluorescent intensity in vacuum and wet-nitrogen atmospheres, respectively, which are described by red-solid lines. The single QDs in both environments exhibited blinking phenomena~\cite{Nirmal_1996}, i.e., an irregular change between the ON and OFF states. The blinking occurrence rate appeared to be reduced in the wet-nitrogen atmosphere, as demonstrated qualitatively in Ref.~\cite{ODA_2007}. It is known that the OFF state of blinking corresponds to the trapping of photogenerated electrons (or holes) in trap site(s) located on or near QD surfaces~\cite{Kuno2001,Nirmal_1996}. Thus, we can qualitatively infer that suppression in the wet-nitrogen atmosphere is due to inactivation of the trap site(s) by photo-adsorption of water molecules in the environment~\cite{ODA_2007}. However, it is difficult to quantify blinking properties with commonly-used conventional analysis, assuming an artificial threshold between the ON and OFF states (Sec.~\ref{Basic_idea_on_State_identification}), due to the low signal-to-noise ratio of the data in Figs.~\ref{implementation_of_experimental_data_in_vac} and \ref{implementation_of_experimental_data_in_wet}. As such, we can apply an HMM analysis described in Secs.~\ref{Hidden Markov model} and \ref{Sec_analysis_for_result_of_HMM} to quantify blinking within the noisy data.

\subsubsection{HMM analysis} 
Green-dashed lines in Figs.~\ref{implementation_of_experimental_data_in_vac} and \ref{implementation_of_experimental_data_in_wet} show time series of the hidden variable $\mathbf{s}^{{\rm ON}}$ in Eq.~(\ref{sON_vec}), which are obtained from the HMM simulations for the time series of the experimental fluorescent intensity $\bm{I}$ in Eq.~(\ref{intensity}). We see from the figure that the behavior of the $\mathbf{s}^{{\rm ON}}$ is in a good agreement with the variation in the experimental intensities (red-solid lines). Also, the $\mathbf{s}^{{\rm ON}}$ basically takes 1 or 0 discontinuously and is hardly affected by noise. For such data, the duration can be estimated stably by the method described in Sec.~\ref{Sec_analysis_for_result_of_HMM}. The blue-dotted line $I_{th}$ is a threshold to distinguish the ON and OFF states for the ${\bm I}$ data, while $s_{th}$ denoted by the black-solid line is a threshold for the $\mathbf{s}^{{\rm ON}}$ data. When estimating the duration in the conventional way with using $I_{th}$ (Sec.~\ref{Basic_idea_on_State_identification}), it becomes erroneous. For example, in the case of Fig.~\ref{implementation_of_experimental_data_in_vac}~(c), the difference in the experimental intensity of the ON and OFF states is clearly smaller than the noise. In this case, it would be difficult to distinguish the two states from fluorescent time series with $I_{th}$. If the way is forcibly performed for estimating the duration, a large amount of short duration due to noise will occur. In contrast, in the analysis for $\mathbf{s}^{{\rm ON}}$ with using $s_{th}$, since the noise is suppressed in $\mathbf{s}^{{\rm ON}}$, the ON/OFF assignment can be stably performed. 

\begin{figure}[!]
\centering
\includegraphics[width=\linewidth]{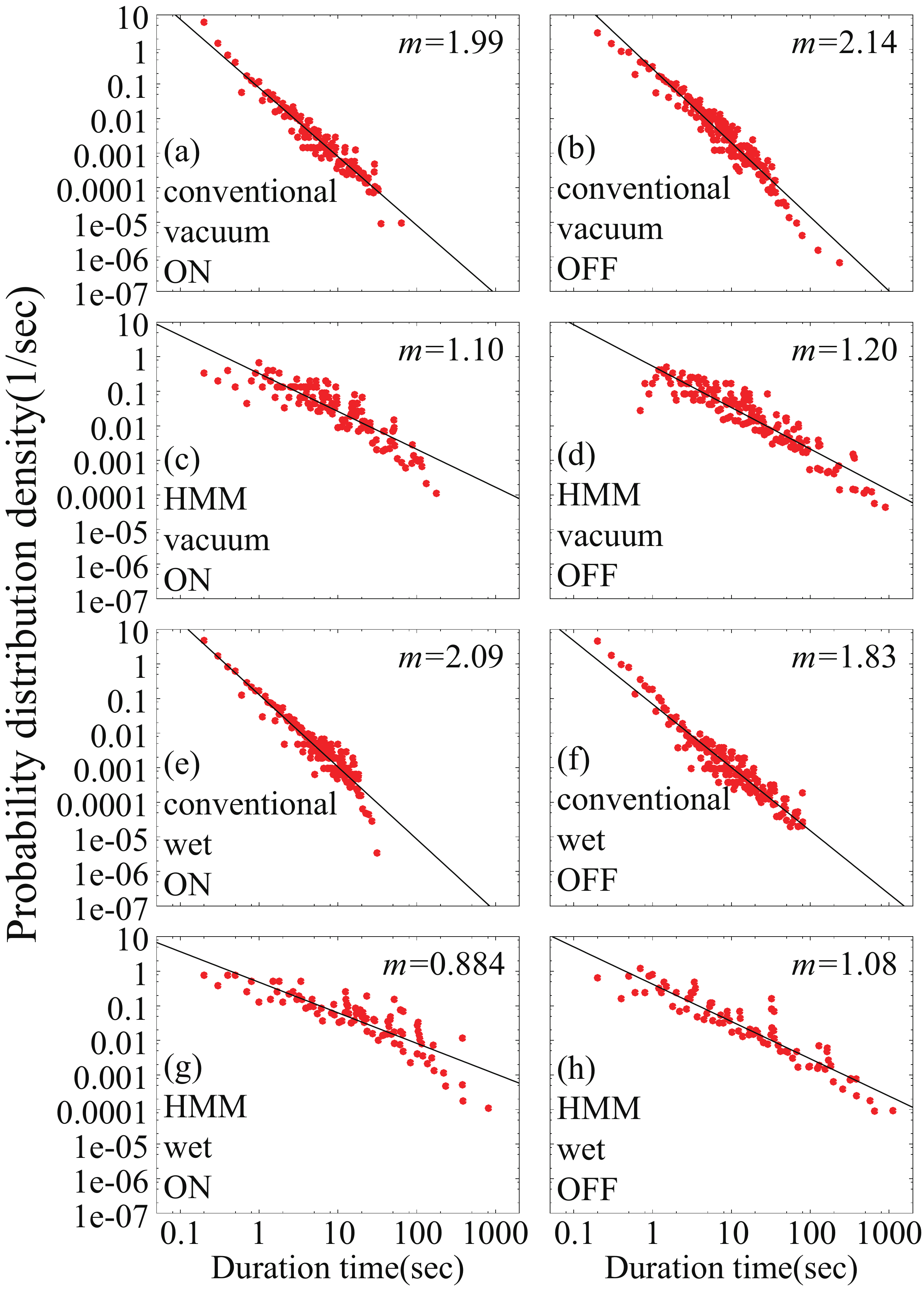}
\caption{Our calculated probability distributions for the ON/OFF duration in the experimental fluorescent time series, $P_{{\rm ON}}(\tau)$ in Eq.~(\ref{dens_ON}) and $P_{{\rm OFF}}(\tau)$ in Eq.~(\ref{dens_OFF}), denoted by red dots. Black-solid lines are $f(\tau)$ in Eq.~(\ref{fitting_function}) and the inverse exponent $m$ represents a slope of the line, whose result is summarized in TABLE~\ref{table_value_of_m}. Also, the small slope corresponds to the high frequency of long duration. Left-side panels (a), (c), (e), and (g) are the results of the ON duration, and the right-side panels (b), (d), (f), and (h) are the results of the OFF duration. Also, the upper four panels [(a), (b), (c), and (d)] show the results for the vacuum condition, and the lower ones [(e), (f), (g), and (h)] show the results under the wet conditions. The (a), (b), (e), and (f) contain the results based on the conventional method , while (c), (d), (g), and (h) represent the results based on the HMM.}
\label{experimental_data_histogram}
\end{figure}

Figure~\ref{experimental_data_histogram} is our calculated probability distributions for the ON/OFF duration, $P_{{\rm ON}}(\tau)$ in Eq.~(\ref{dens_ON}) and $P_{{\rm OFF}}(\tau)$ in Eq.~(\ref{dens_OFF}), denoted by red dots. Left-side panels (a), (c), (e), and (g) are the results of the ON duration, and the right-side panels (b), (d), (f), and (h) are the results of the OFF duration. Also, the upper four panels [(a), (b), (c), and (d)] show the results for the vacuum condition, and the lower ones [(e), (f), (g), and (h)] show the results under the wet conditions. The (a), (b), (e), and (f) panels contain the results based on the conventional method, while the (c), (d), (g), and (h) ones represent the results based on the HMM. To analyze trends of each data, we performed a fitting  of the following function~\cite{Shimizu_2001} to the data as 
\begin{eqnarray}
f(\tau)= A \tau ^{-m}, \label{fitting_function}
\end{eqnarray}
where $A$ is a coefficient and $m$ is an inverse exponent. The fitted function is described by a black-solid line and the obtained $m$ values are summarized in TABLE~\ref{table_value_of_m}. A small $m$ indicates that long-duration states tend to be formed; for example, the plot of the panel~(g) exhibits the smallest $m=0.884$ and therefore has much long duration data compared with the other plots. In contrast, in the largest $m$ plot of the panel~(b), the data concentrate in the short duration regime. By comparison of $m$ of each plot, we found that the probability distributions based on the conventional method are approximated by $\tau ^{-2}$, while the probability distributions based on the HMM decay as $\tau ^{-1}$. Thus, the plots based on the conventional method tend clearly to reflect many artificial short duration states, and the HMM corrects the long duration data. We note that, on the results based on the HMM, the $m$ values with the wet condition are basically small compared to the $m$ with the vacuum condition, indicating that single QDs emit long and quench long in the wet atmospheres, while in the vacuum atmosphere, the single QDs blink with moderate length.
\begin{table}[!]
\caption{A summary of the fitting of $f(\tau)$ [Eq.~(\ref{fitting_function})] to the data of Fig.~\ref{experimental_data_histogram}, where we list the inverse exponent $m$ and its standard errors.} 
\begin{center}
\begin{tabular}{c@{\ \ }c@{\ \ }c@{\ \ }c@{\ \ }c} 
\hline \hline \\ [-8pt] 
  &  \multicolumn{2}{c}{Vacuum} & \multicolumn{2}{c}{Wet} \\ [2pt] 
 \hline \\ [-8pt]
 & ON & OFF & ON & OFF \\ [2pt] 
 \hline \\ [-8pt]
Conventional & 1.99$\pm$0.05 & 2.14$\pm$0.04 & 2.09$\pm$0.06 & 1.83$\pm$0.04  \\ [2pt] 
HMM          & 1.10$\pm$0.05 & 1.20$\pm$0.04 & 0.88$\pm$0.06 & 1.08$\pm$0.05  \\ [2pt] 
\hline \hline
\end{tabular} 
\end{center}
\label{table_value_of_m}
\end{table}

\subsection{HMM analysis for theoretical benchmark data} \label{Sec_implementation_of_theoretical_data}
In this part, we check quantitative accuracy of the present HMM simulation. In Sec.~\ref{Sec_implementation_of_experimental_data}, we have shown that there is a discernible difference between the results based on the conventional method and the HMM simulation, but it does not mean the quantitative accuracy for the HMM. In order to verify the accuracy of the HMM, it is necessary to perform HMM analyses for a time series with correct answers and check whether the HMM can reproduce the correct results.

\subsubsection{Generation of model time series}  
For this purpose, we consider the following model distribution functions for the ON and OFF duration~\cite{Bruhn2014} as  
\begin{eqnarray}
 p_{{\rm ON}}(\tau) = A^{{\rm ON}} \tau^{-q} e^{-\tau/\xi}
 \theta(\tau-\tau_{min}) \theta(\tau_{max}-\tau) 
 \label{PON_model} 
 \end{eqnarray}
 and 
 \begin{eqnarray} 
 p_{{\rm OFF}}(\tau) = A^{{\rm OFF}} \tau^{-l}
 \theta(\tau-\tau_{min}) \theta(\tau_{max}-\tau), 
 \label{POFF_model} 
\end{eqnarray}
respectively, where $A^{{\rm ON}}$ and $A^{{\rm OFF}}$ are normalization constants. $\theta(x)$ is a step function, and $q$, $\xi$ and $l$ are parameters of the model functions. $\tau_{min}$ and $\tau_{max}$ are respectively lower and upper cutoffs of the duration considered in the present model. These functions have been widely used in analyses for the blinking phenomena with the power law of QDs~\cite{Bruhn2014,kevin_2013}. The exponential term in Eq.~(\ref{PON_model}) is valid in the case of the truncated power law. In the simulation, we first generate a time series with the model distribution functions. Next, we perform an HMM analysis for the generated time series, and evaluate the ON and OFF duration. Finally, we check whether the distribution function calculated with the simulation [Eqs.~(\ref{dens_ON}) and (\ref{dens_OFF})] reproduces the original model distribution functions [Eqs.~(\ref{PON_model}) and (\ref{POFF_model})].

The model time series is generated as follows: 
\begin{enumerate}
    \item We first define baselines for the ON and OFF intensities ($I_{{\rm ON}}$ and $I_{{\rm OFF}}$) and their difference 
      \begin{eqnarray}
        \Delta = I_{{\rm ON}} - I_{{\rm OFF}}. 
        \label{difference_ION_IOFF} 
      \end{eqnarray}
    \item We next sample \{$\tau^{{\rm ON}}_{\alpha}$\} from $p_{{\rm ON}}(\tau)$ in Eq.~(\ref{PON_model}) and \{$\tau^{{\rm OFF}}_{\alpha}$\} from $p_{{\rm OFF}}(\tau)$ in Eq.~(\ref{POFF_model}) with $\alpha$ specifying a sampling number. The $\tau_{\alpha}^{{\rm ON}}$ and $\tau_{\alpha}^{{\rm OFF}}$ represent the duration of the $I_{{\rm ON}}$ and $I_{{\rm OFF}}$ intensities, respectively. With these data, we make a sequence $\bm{\tau}$ as 
    \begin{eqnarray}
     \bm{\tau}=(\tau^{{\rm ON}}_1, \tau^{{\rm OFF}}_1, \tau^{{\rm ON}}_2, \cdots, 
     \tau^{{\rm ON}}_{N_e}, \tau^{{\rm OFF}}_{N_e}), 
     \label{tau_for_I0} 
    \end{eqnarray}
    where $N_e$ is the total number of the ON or OFF events. 
    \item Then, based on the above $\bm{\tau}$, $I_{{\rm ON}}$, and $I_{{\rm OFF}}$, we construct a bare time series  
    \begin{eqnarray}
     {\bm I}_0=(I^0_1, I^0_2, \cdots, I^0_N)^T  
     \label{intensity_0} 
    \end{eqnarray}
    which are discretized at interval $\Delta t$ and in a similar form to Eq.~(\ref{intensity}). 
    \item Finally, we add an artificial noise $\eta$ to the above time series $\bm{I}_0$, where $\eta$ is sampled from the Gaussian functions as 
    \begin{eqnarray}
      \eta \sim \mathcal{N}(\eta|0,\sigma ^2) =\frac{1}{\sqrt{2\pi\sigma^2}}\exp\biggl({-\frac{\eta^2}{2\sigma^2}}\biggr) 
      \label{noise_eta}
    \end{eqnarray}
    with $\sigma$ being a standard deviation of the Gaussian function. Thus, the final model time series $\bm{I}$ in Eq.~(\ref{intensity}) is obtained. 
\end{enumerate}

\begin{figure*}[!]
\centering
\includegraphics[width=1.0\linewidth]{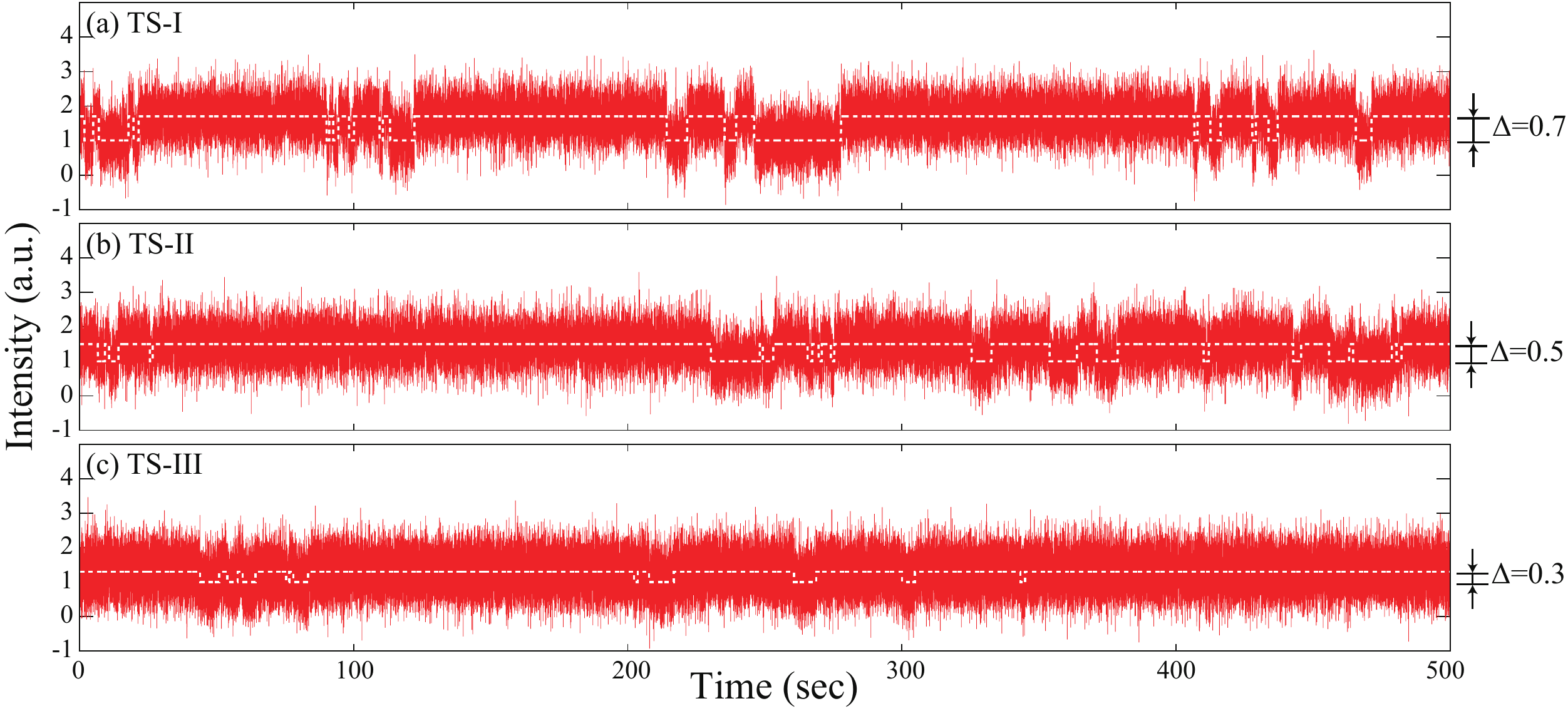}
\vspace{-0.3cm} 
\caption{Model time series ${\bm I}$ [Eq.~(\ref{intensity})] generated by a manner in Sec.~\ref{Sec_implementation_of_theoretical_data}, denoted by red-solid lines. White-dashed lines are the bare time series $\bm{I}_0$ in Eq.~(\ref{intensity_0}). $\Delta=I_{{\rm ON}}-I_{{\rm OFF}}$ in Eq.~(\ref{difference_ION_IOFF}) represents the bare intensity difference. (a) TS-I ($\Delta=0.7$), (b) TS-II ($\Delta=0.5$), and (c) TS-III ($\Delta=0.3$). The noise $\eta$ is 0.5. The time-grid interval $\Delta t$ was set to 0.01 s for TS-I and TS-II and 0.005 s for TS-III and a single time series contains 20-ON and 20-OFF events ($N_e$ = 20).}
\label{photon_signal_time-series_data}
\end{figure*}

Figure~\ref{photon_signal_time-series_data} shows the generated time series $\bm{I}$ in Eq.~(\ref{intensity}) (Red-solid lines). The bare time series $\bm{I}_0$ in Eq.~(\ref{intensity_0}) is also shown with white-dashed lines. Parameters in generating the time series and characterizing the model functions $p_{{\rm ON}}(\tau)$ in Eq.~(\ref{PON_model}) and $p_{{\rm OFF}}(\tau)$ in Eq.~(\ref{POFF_model}) are summarized in TABLE~\ref{table_jikeiretsu}. In the present study, we consider three time series with different $\Delta$ (0.7, 0.5, and 0.3) in Eq.~(\ref{difference_ION_IOFF}) with the noise amplitude $\eta$ in Eq.~(\ref{noise_eta}) kept at 0.5. We call the time series with $\Delta = 0.7$, 0.5, and 0.3 TS-I, TS-II, and TS-III, respectively. The time-grid interval $\Delta t$ of the time series was set to 0.01 s for TS-I and TS-II and 0.005 s for TS-III and one time series contains 20-ON and 20-OFF events ($N_e$ = 20). We generated 1000 samples for each time series and took an ensemble average of them in the analysis. The panels (a), (b), and (c) in Fig.~\ref{photon_signal_time-series_data} compare TS-I, TS-II, and TS-III. As the $\Delta$ becomes smaller ($\Delta=0.7 \to 0.5 \to 0.3$), it becomes difficult to distinguish between the ON and OFF intensities. At $\Delta$ = 0.3 (TS-III), it would be no longer possible for the human eye to distinguish between the ON and OFF states correctly.
\begin{figure*}[!]
\centering
\includegraphics[width=0.985\linewidth]{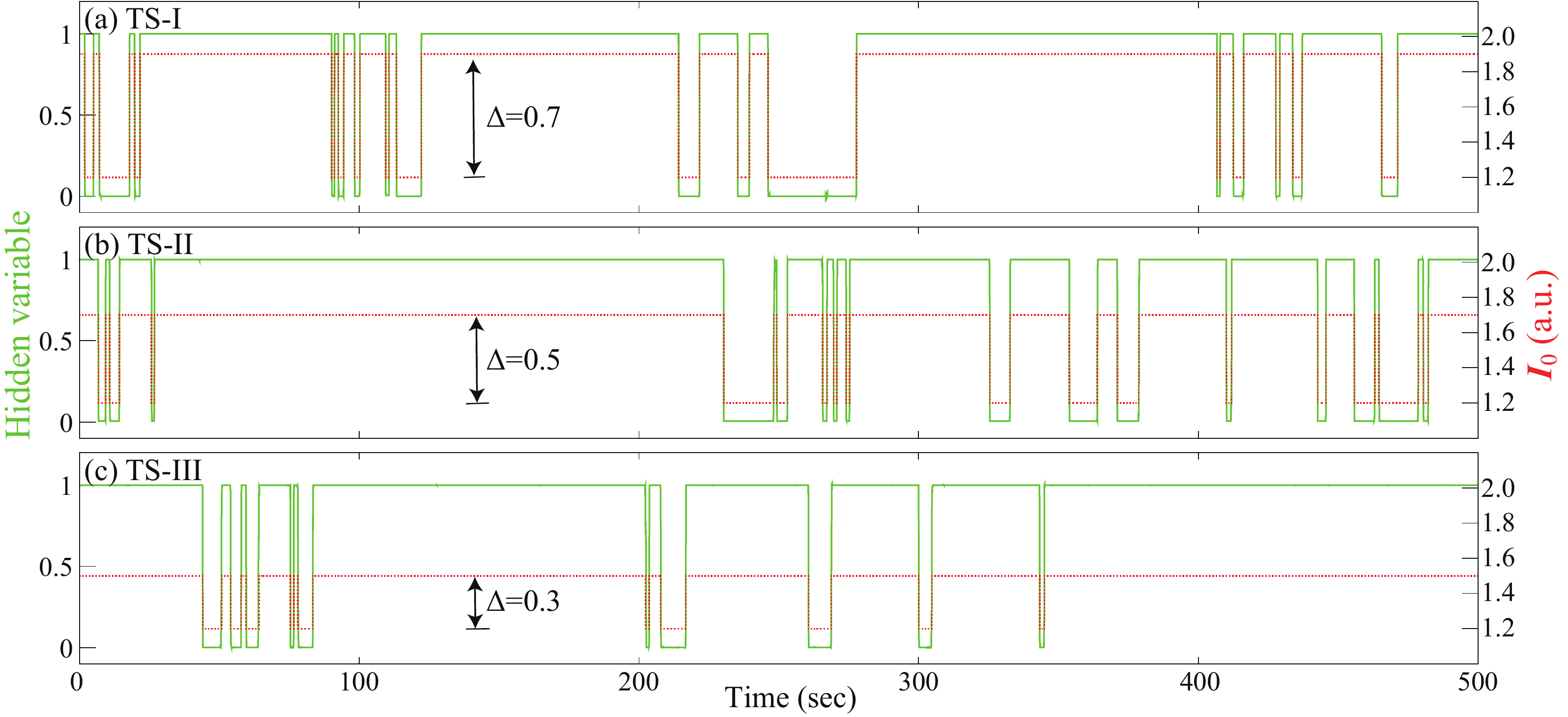}
\vspace{-0.3cm} 
\caption{Calculated time series of hidden variable $\mathbf{s}^{{\rm ON}}$ [Eq.~(\ref{sON_vec})] with the HMM simulations, denoted by green-solid lines. The panel (a) shows the $\mathbf{s}^{{\rm ON}}$ obtained for TS-I ($\Delta=0.7$) in Fig.~\ref{photon_signal_time-series_data} (a). Similarly, the panels (b) and (c) are the results for TS-II ($\Delta=0.5$) in Fig.~\ref{photon_signal_time-series_data} (b) and TS-III ($\Delta=0.3$) in Fig.~\ref{photon_signal_time-series_data} (c), respectively. Bare time series ${\bm I}_0$ denoted by red-dotted lines are also shown for comparison. If the HMM simulation is successful, the $\mathbf{s}^{{\rm ON}}$ should match ${\bm I}_0$.}
\label{Hidden_variable}
\end{figure*}
\begin{table}[!]
\caption{Parameters for generating model time series of Fig.~\ref{photon_signal_time-series_data}, based on the model functions $p_{{\rm ON}}(\tau)$ in Eq.~(\ref{PON_model}) and $p_{{\rm OFF}}(\tau)$ in Eq.~(\ref{POFF_model}). $\xi$, $q$, $l$, $\tau_{max}$, and $\tau_{min}$ are parameters characterizing $p_{{\rm ON}}(\tau)$ and $p_{{\rm OFF}}(\tau)$. $I_{{\rm ON}}$ and $I_{{\rm OFF}}$ are baselines for the ON and OFF intensities, respectively, and $\Delta$ is their difference [Eq.~(\ref{difference_ION_IOFF})]. According to the $\Delta$, we named the time series with $\Delta = 0.7$, 0.5, and 0.3 TS-I, TS-II, and TS-III, respectively. $\sigma$ is a standard deviation of the Gaussian function to generate noise $\eta$ [Eq.~(\ref{noise_eta})]. $N_e$ is the total number of the ON or OFF events [Eq.~(\ref{tau_for_I0})] in the time series. Note that the total number of the ON and OFF events in the time series is the same. $\Delta t$ is a grid spacing of time series, and $\Delta \tau$ is a grid spacing of the duration grid, which is introduced in the calculation of $P_{{\rm ON}}(\tau)$ in Eq.~(\ref{dens_ON}) and $P_{{\rm OFF}}(\tau)$ in Eq.~(\ref{dens_OFF}). The $\tau_{max}$, $\tau_{min}$, $\Delta \tau$, and $\Delta t$ are given in a unit of s.} 
\begin{center}
\begin{tabular}{l@{\ \ \ \ }c@{\ \ \ \ }c@{\ \ \ \ }c@{\ \ \ \ }c} 
\hline \hline \\ [-8pt] 
 & Eq. & TS-I & TS-II & TS-III \\ [2pt] 
 \hline \\ [-8pt]
$\xi$ & (\ref{PON_model})  & 1800 & 1800 & 1800 \\ [2pt] 
$q$   & (\ref{PON_model})  & 1.3  & 1.3  & 1.3  \\ [2pt] 
$l$   & (\ref{POFF_model}) & 1.7  & 1.7  & 1.7  \\ [2pt] 
$\tau_{max}$ & (\ref{PON_model})(\ref{POFF_model}) & 1000 & 1000 & 500 \\ [2pt] 
$\tau_{min}$ & (\ref{PON_model})(\ref{POFF_model}) & 1    & 1    & 1   \\ [2pt] 
$\Delta \tau$ & (\ref{condition_delta_t_and_tau}) & 0.1 & 0.1 & 0.1 \\ [2pt] 
\hline \\ [-8pt] 
$I_{{\rm ON}}$  & (\ref{difference_ION_IOFF})  & 1.7 & 1.5 & 1.3 \\ [2pt] 
$I_{{\rm OFF}}$ & (\ref{difference_ION_IOFF})  & 1.0 & 1.0 & 1.0 \\ [2pt] 
$\Delta$ & (\ref{difference_ION_IOFF})         & 0.7 & 0.5 & 0.3 \\ [2pt] 
$\sigma$ & (\ref{noise_eta})  & 0.5 & 0.5 & 0.5 \\ [2pt] 
$N_{e}$  & (\ref{tau_for_I0}) & 20  & 20  & 20  \\ [2pt] 
$N_{itr}$ & (\ref{s_i_max}) & 200  & 200  & 1000 \\ [2pt]  
$\Delta t$ & (\ref{condition_delta_t_and_tau}) & 0.01 & 0.01 & 0.005 \\ [2pt]  
$N_{sample}$ & - & 1000 & 1000 & 1000 \\ [2pt]  
\hline \hline
\end{tabular} 
\end{center}
\label{table_jikeiretsu}
\end{table}

We note that, to ensure numerical accuracy of the HMM simulation, a proper choice of a grid spacing $\Delta t$ of the time series is important; $\Delta t$ must ideally be small enough compared to a grid spacing $\Delta \tau$ of the duration grid which is introduced in the calculation of $P_{{\rm ON}}(\tau)$ in Eq.~(\ref{dens_ON}) and $P_{{\rm OFF}}$ in Eq.~(\ref{dens_OFF}). Then,  
\begin{eqnarray}
    \Delta t \ll \Delta \tau.  
    \label{condition_delta_t_and_tau}
\end{eqnarray}
By definition, $\Delta \tau$ represents the minimum duration. In order to evaluate it accurately, the $\Delta t$ must be small enough. In the present study, $\Delta t$ and $\Delta \tau$ were respectively set to 0.01 s and 0.1 s for TS-I and TS-II and 0.005 s and 0.1 s for TS-III. 

\subsubsection{The time series of hidden variables}\label{Time-series_data_of_hidden_variable}
We show in Fig.~\ref{Hidden_variable} results of our HMM analysis for the time series in Fig.~\ref{photon_signal_time-series_data}. The panels (a), (b), and (c) correspond to the results for TS-I [Fig.~\ref{photon_signal_time-series_data}~(a)], TS-II [Fig.~\ref{photon_signal_time-series_data}~(b)], and TS-III [Fig.~\ref{photon_signal_time-series_data}~(c)], respectively. The solid-green line describes the time series of the hidden variables $\mathbf{s}^{{\rm ON}}$ in Eq.~(\ref{sON_vec}), which is compared with the bare time series $\bm{I}_0$ in Eq.~(\ref{intensity_0}) (red-dotted line). We see that the behavior of $\mathbf{s}^{{\rm ON}}$ rather well reproduces the $\bm{I}_0$. There are almost no misjudgements even for the time series with $\Delta=0.3$ (TS-III), and thus our HMM has very high accuracy.

\subsubsection{Probability density of duration time}\label{Sec_Probability_density_of_duration_time}
\begin{figure*}[!]
\centering
\includegraphics[width=0.7\linewidth]{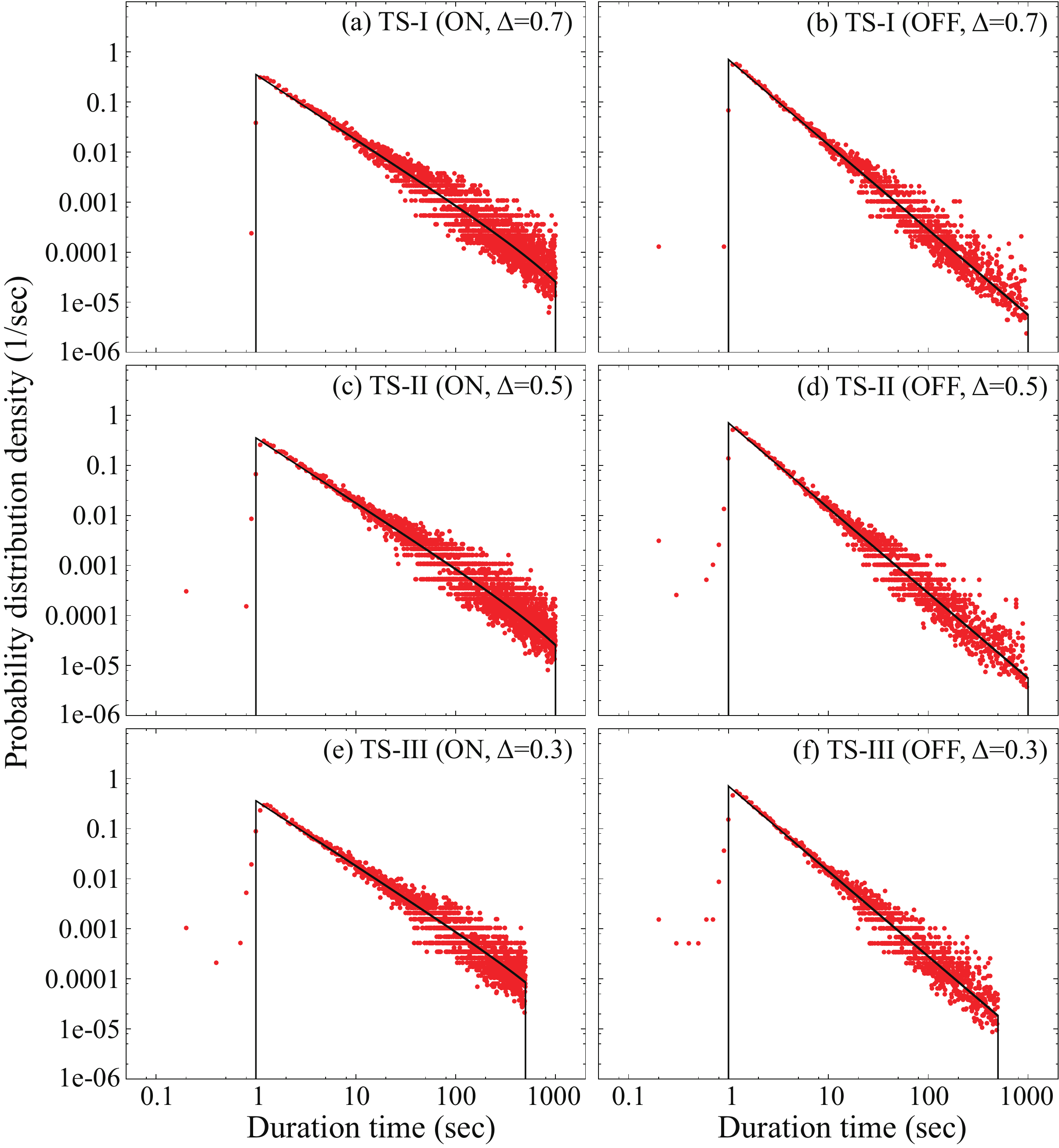}
\caption{Our calculated probability distributions for the ON/OFF duration in the fluorescent time series, $P_{{\rm ON}}(\tau)$ in  Eq.~(\ref{dens_ON}) and $P_{{\rm OFF}}(\tau)$ in Eq.~(\ref{dens_OFF}), denoted by red dots. Black-solid lines represent model probability distributions $p_{{\rm ON}}(\tau)$ in Eq.~(\ref{PON_model}) and $p_{{\rm OFF}}(\tau)$ in Eq.~(\ref{POFF_model}). The upper (a) and (b), middle (c) and (d), and lower (e) and (f) panels correspond to the results for the time series with $\Delta=0.7$, 0.5, and 0.3, respectively. Note that each result is obtained for $N_{sample}$ time series. Also, left (a), (c), and (e) panels are the results for the ON duration, while right (b), (d), and (f) panels are the results of the OFF state.} 
\label{Probability_density_of_duration_time}
\end{figure*}

From the analysis to the $\mathbf{s}^{{\rm ON}}$ time series as shown in Fig.~\ref{Hidden_variable}, we evaluated the duration \{$\tau_{\alpha}^{{\rm ON}}$\} and \{$\tau_{\alpha}^{{\rm OFF}}$\} using the method described in Sec.~\ref{Sec_analysis_for_result_of_HMM}, and calculated their probability distributions $P_{{\rm ON}}(\tau)$ in Eq.~(\ref{dens_ON}) and $P_{{\rm OFF}}(\tau)$ in Eq.~(\ref{dens_OFF}). The resulting blinking plots are shown in Fig.~\ref{Probability_density_of_duration_time} by red dots. The left-side panels represent $P_{{\rm ON}}(\tau)$, and the right-side panels are $P_{{\rm OFF}}(\tau)$. Also, the upper (a) and (b) panels describe the results for the model time series TS-I [Fig.~\ref{photon_signal_time-series_data} (a)]. Similarly, the middle (c) and (d) and lower (e) and (f) panels describe the results for the time series TS-II [Fig.~\ref{photon_signal_time-series_data} (b)] and TS-III [Fig.~\ref{photon_signal_time-series_data} (c)], respectively. We also give the original model distributions $p_{{\rm ON}}(\tau)$ in Eq.~(\ref{PON_model}) and $p_{{\rm OFF}}(\tau)$ in Eq.~(\ref{POFF_model}) by black-solid lines. We see that the calculated blinking plots are well reproduce the original distribution functions. As emphasized in Fig.~\ref{photon_signal_time-series_data}~(c), it is difficult for humans to identify the ON or OFF state in the $\bm{I}$ time series TS-III ($\Delta=0.3$, $\eta=0.5$), but the present HMM simulation is able to discriminate between the ON and OFF states with a fairly high accuracy.

\subsubsection{Quantitative accuracy of HMM analysis}\label{Quantitative accuracy of HMM analysis}

To check the HMM-simulation accuracy quantitatively, we evaluated recall and precision, which are defined as 
\begin{eqnarray}
 {\rm Recall} = \frac{N_{suc}}{N_{ans}}, 
 \label{recall}
\end{eqnarray}
and
\begin{eqnarray}
 {\rm Precision} = \frac{N_{suc}}{N_{sim}},
 \label{precision}
\end{eqnarray}
respectively. Here, $N_{ans}$, $N_{sim}$, and $N_{suc}$ are the number of actual events with a certain duration, the number of events with a certain duration obtained by the HMM simulation, and the number of correctly predicted events,  respectively. In the present model time series, we introduce cutoffs $\tau_{min}$ and $\tau_{max}$ [Eqs.~(\ref{PON_model}) and (\ref{POFF_model})], and then the actual events exist only in 1-1000 s. In this analysis, we tolerate a 10\% error. For example, in the case of data with the duration of $\tau=1$ s, if the simulation evaluates it within 0.9-1.1 s, it is regarded as a success event. TABLE~\ref{simulation_accuracy} shows analysis results for the HMM simulation, where we accumulated the duration data from the $N_{sample}$ time series, where $N_{sample}$ is given in TABLE~\ref{table_jikeiretsu}. In the table, we present the duration results digit by digit. From the table, we see that, for all the time series, accuracy of the recall and precision are satisfactory. Certainly the accuracy of the short duration seems a little bad; especially, simulations observes duration of 0-1 s that does not actually exist. Overall, however, the accuracy is very high and satisfactory. 
\begin{table*}[!]
    \centering
    \caption{Analysis of the HMM prediction accuracy. Upper, middle, and lower blocks describe the analysis results for the time series with $\Delta=0.7$, 0.5, and 0.3, respectively. Note that each result is obtained for $N_{sample}$ time series. In the table, we present the duration results digit by digit. $N_{ans}$ is the total number of actual events in the ON or OFF states, which is evaluated from ${\bm I}_0$ in Eq.~(\ref{intensity_0}). $N_{sim}$ is the total number of events in the ON or OFF states, which is estimated from the time series ${\bf s}^{{\rm ON}}$ in Eq.~(\ref{sON_vec}), based on the HMM simulation. $N_{suc}$ is the total number of correctly predicted events. Recall and precision are $N_{suc}/N_{ans}$ in Eq.~(\ref{recall}) and $N_{suc}/N_{sim}$ in Eq.~(\ref{precision}), respectively. These results are for  $N_{sample}$ time series in TABLE~\ref{table_jikeiretsu}. The unit of $\tau$ is sec.}
    \begin{tabular}{c@{\ \ }c@{\ \ }c@{\ \ }c@{\ \ }c@{\ \ }c@{\ \ \ \ \ \ }c@{\ \ }c@{\ \ }c@{\ \ }c@{\ \ }c@{\ \ }c@{\ \ }}
      \hline \hline \\ [-5pt] 
      TS-I & \multicolumn{5}{c}{ON} & & \multicolumn{5}{c}{OFF} \\ [0pt] 
      \hline \\ [-5pt] 
      $\tau$ & $N_{ans}$ & $N_{sim}$ & $N_{suc}$ & Recall & Precision & 
             & $N_{ans}$ & $N_{sim}$ & $N_{suc}$ & Recall & Precision \\ [2pt] \hline \\ [-5pt] 
    0-1 & 0         & 358         & 0         & -                 & 0                 & 
             & 0         & 363         & 0         & -                 & 0 \\ [5pt] 
      1-10   & 11832     & 11804     & 11625     & 0.983               & 0.985                 & 
             & 16058     & 16025     & 15682     & 0.977               & 0.979 \\ [5pt] 
      10-100 & 5728      & 5728      & 5728      & 1.0               & 1.0                 & 
             & 3325      & 3325      & 3325      & 1.0               & 1.0 \\ [5pt] 
    100-1000 & 2440      & 2440      & 2440      & 1.0               & 1.0                 & 
             & 617       & 617       & 617       & 1.0               & 1.0 \\ [5pt] 
      1000-  & 0         & 0         & 0         & -                 & -                 & 
             & 0         & 0         & 0         & -                 & - \\ [2pt] 
      \hline \hline \\ [-5pt] 
      TS-II & \multicolumn{5}{c}{ON} & & \multicolumn{5}{c}{OFF} \\ [0pt] 
      \hline  \\ [-5pt] 
      $\tau$ & $N_{ans}$ & $N_{sim}$ & $N_{suc}$ & Recall & Precision & 
             & $N_{ans}$ & $N_{sim}$ & $N_{suc}$ & Recall & Precision \\ [2pt] \hline \\ [-5pt]
    0-1 & 0         & 1396         & 0         & -                 & 0                 & 
             & 0         & 1459         & 0         & -                 & 0 \\ [5pt]
      1-10   & 11702     & 11620     & 10747     & 0.918             & 0.925                 & 
             & 16123     & 15978     & 14501     & 0.899             & 0.908 \\ [5pt] 
    10-100   & 5844      & 5844      & 5844      & 1.0               & 1.0                 & 
             & 3247      & 3247      & 3247      & 1.0               & 1.0 \\ [5pt] 
    100-1000 & 2454      & 2454      & 2454      & 1.0               & 1.0                 &
             & 630       & 630       & 630       & 1.0               & 1.0 \\ [5pt] 
    1000-    & 0         & 0         & 0         & -                 & -                 & 
             & 0         & 0         & 0         & -                 & - \\ [2pt]  
    \hline \hline \\ [-5pt] 
      TS-III & \multicolumn{5}{c}{ON} & & \multicolumn{5}{c}{OFF} \\ [0pt] 
      \hline \\ [-5pt] 
     $\tau$ & $N_{ans}$ & $N_{sim}$ & $N_{suc}$ & Recall & Precision & 
            & $N_{ans}$ & $N_{sim}$ & $N_{suc}$ & Recall & Precision \\ [2pt] \hline \\ [-5pt]
   0-1 & 0         & 2031         & 0         & -                 & 0                 & 
             & 0         & 2163         & 0         & -                 & 0 \\ [5pt] 
      1-10  & 12085     & 11926      & 10398      & 0.86             & 0.872                  & 
            & 16236     & 15945     & 13458     & 0.829             & 0.844 \\ [5pt] 
     10-100 & 5933      & 5934      & 5931      & 0.999             & 0.999                  & 
            & 3210      & 3212      & 3208      & 0.999             & 0.999 \\ [5pt] 
   100-500 & 1982      & 1982      & 1981      & 0.999             & 0.999                  & 
            & 554       & 554       & 554       & 1.0             & 1.0 \\ [5pt] 
      500- & 0         & 0         & 0         & -                 & -                  & 
            & 0         & 0         & 0         & -                 & - \\ [2pt] 
      \hline \hline
    \end{tabular}
    \label{simulation_accuracy}
\end{table*}

\section{Summary}\label{sec_summary} 
We have presented an HMM analysis for experimental and theoretical fluorescent time series of QDs. In this simulation, we have calculated the time series of hidden variables to evaluate the ON/OFF duration of the fluorescence. With the resulting duration data, we have calculated blinking plots. Through the comparison between the results based on the conventional and HMM analyses, we have found discernible quantitative differences; in the conventional method, the ON or OFF state is directly evaluated for the noisy fluorescent time series, thus leading to a large amount of artificial short duration data. On the other hand, in the HMM analysis, the hidden-variable time series which is noise-suppressed is calculated for the ON/OFF assignment, so we can accumulate reliable duration data; the artificial short duration data are suppressed and long duration data are properly evaluated. It was found that these differences in the evaluation methods have a great influence on the analysis for the noisy time-series data;  in the case of the QD fluorescence data, it had a significant effect on the power of the duration probability distribution. Also, in order to show the accuracy of the present HMM analysis, we have analyzed the theoretical benchmark time series generated from model distribution functions for the duration. The distribution functions obtained from the HMM simulations well reproduce the model distribution functions, and we have found that the ON/OFF assignments can be performed rather accurately even for the low signal-to-noise time series.

In the present simulation, we have focused on two-state analysis, but the HMM can also be applied to analysis of three or more states. If such an analysis is performed successfully, it will be possible to construct a more detailed model on fluorescence of QDs, especially on the relaxation from photo-excited to ground states~\cite{Hou_2019}, which is left as a future task. 

\section{Acknowledgments}\label{acknowledgments}
We thank Yoshihide Yoshimoto for helpful discussions. We acknowledge the financial support of JSPS Kakenhi Grant No. 16H06345,  No. 16K05452, No. 17H03393, No. 17H03379, No. 19K03673, and 22H01183. A part of the computation was done at Supercomputer Center, Institute for Solid State Physics, University of Tokyo. 


\appendix
\section{Blocking Gibbs sampling} \label{Blocking Gibbs sampling}
In this appendix, we describe how to calculate the conditional distribution function $p(\mathbf{S},{\bm \mu}, {\bm \lambda}, \bm{\pi}, \mathbf{A}|\bm{I})$ in Eq.~(\ref{conditional-p}). For this purpose, we use the Gibbs sampling~\cite{ZOUBIN_2001,Stuart_1984} which gives the approximate solution of $p(\mathbf{S},{\bm \mu}, {\bm \lambda}, \bm{\pi}, \mathbf{A}|\bm{I})$. The calculation proceeds as follows: 
\begin{enumerate}
 \item We first calculate a conditional probability distribution of $p(\mathbf{S}^{(i)}|\bm{\mu}^{(i-1)},\bm{\lambda}^{(i-1)},\bm{\pi}^{(i-1)},\mathbf{A}^{(i-1)}, \bm{I})$ with fixed by $\bm{\mu}^{(i-1)}$, $\bm{\lambda}^{(i-1)}$, $\bm{\pi}^{(i-1)}$, $\mathbf{A}^{(i-1)}$, and $\bm{I}$, and sample $\mathbf{S}^{(i)}$ from the resulting conditional probability distribution as
   \begin{eqnarray}
    \mathbf{S}^{(i)} 
    \sim 
    p(\mathbf{S}^{(i)}|\bm{\mu}^{(i-1)},\bm{\lambda}^{(i-1)},\bm{\pi}^{(i-1)},\mathbf{A}^{(i-1)},\bm{I}),  
    \label{gibbs_s} 
   \end{eqnarray}
 where the upper suffix $i$ specifies the iteration step of the Gibbs sampling, and the variables of the initial step, $\bm{\mu}^{(0)}$, $\bm{\lambda}^{(0)}$, $\bm{\pi}^{(0)}$, $\mathbf{A}^{(0)}$ are evaluated with the hyperparameters in TABLE~\ref{table-HMM}.

 \item We next calculate a conditional probability distribution $p(\bm{\mu}^{(i)},\bm{\lambda}^{(i)}|\mathbf{S}^{(i)},\bm{\pi}^{(i-1)},\mathbf{A}^{(i-1)}, \bm{I})$ with fixed $\mathbf{S}^{(i)}$, $\bm{\pi}^{(i-1)}$, $\mathbf{A}^{(i-1)}$, and $\bm{I}$, and sample $\bm{\mu}^{(i)}$, $\bm{\lambda}^{(i)}$ from the obtained conditional distribution function as 
 \begin{eqnarray}
   \bm{\mu}^{(i)},\bm{\lambda}^{(i)}
   \sim 
   p(\bm{\mu}^{(i)},\bm{\lambda}^{(i)}|\mathbf{S}^{(i)},\bm{\pi}^{(i-1)},\mathbf{A}^{(i-1)}, \bm{I}). 
   \label{gibbs_mu_lambda}
  \end{eqnarray}

\item The same treatment applies to $\bm{\pi}^{(i)}$; we evaluate a conditional distribution function $p(\bm{\pi}^{(i)}|\mathbf{S}^{(i)}, \bm{\mu}^{(i)}, \bm{\lambda}^{(i)}, \mathbf{A}^{(i-1)}, \bm{I})$ and sample $\bm{\pi}^{(i)}$ as follows:  
\begin{eqnarray}
   \bm{\pi}^{(i)}
   \sim
   p(\bm{\pi}^{(i)}|\mathbf{S}^{(i)},\bm{\mu}^{(i)},\bm{\lambda}^{(i)},\mathbf{A}^{(i-1)}, \bm{I}).
   \label{gibbs_pi}
  \end{eqnarray}

\item Finally, we sample $\mathbf{A}^{(i)}$ from a conditional distribution function with fixed $\mathbf{S}^{(i)}$, $\bm{\mu}^{(i)}$, $\bm{\lambda}^{(i)}$, $\bm{\pi}^{(i)}$, and $\bm{I}$ as follows:  
   \begin{eqnarray}
   \mathbf{A}^{(i)}
   \sim
   p(\mathbf{A}^{(i)}|\mathbf{S}^{(i)},\bm{\mu}^{(i)},\bm{\lambda}^{(i)},\bm{\pi}^{(i)}, \bm{I}). 
   \label{gibbs_A}
  \end{eqnarray}
\end{enumerate}
One iteration loop of the Gibbs sampling consists of Eq.~(\ref{gibbs_s}) $\to$ Eq.~(\ref{gibbs_mu_lambda}) $\to$ Eq.~(\ref{gibbs_pi}) $\to$ Eq.~(\ref{gibbs_A}) $\to$  Eq.~(\ref{gibbs_s}). The $\mathbf{S}^{(i)}$, $\bm{\mu}^{(i)}$, $\bm{\lambda}^{(i)}$, $\bm{\pi}^{(i)}$, $\mathbf{A}^{(i)}$ are optimized by repetition of this iteration. In the present HMM, since each conditional probability distribution is analytically obtained (see below), the Gibbs sampling is especially effective.

We next describe the conditional distribution for Eqs.~(\ref{gibbs_s}), (\ref{gibbs_mu_lambda}), (\ref{gibbs_pi}), and (\ref{gibbs_A}). The calculation is a little complicated, but the derivation itself is straightforward, so only the results are shown.
\begin{enumerate}
\item $p(\mathbf{S}^{(i)}|\bm{\mu}^{(i-1)}, \bm{\lambda}^{(i-1)}, \bm{\pi}^{(i-1)}, \mathbf{A}^{(i-1)}, \bm{I})$ [Eq.~(\ref{gibbs_s})]:

    We used a blocking Gibbs sampling based on a forward-backward algorithm~\cite{ZOUBIN_2001,RABINER_1989,Baum_1972}. In this treatment, we utilize a marginalization as 
    \begin{eqnarray}
      p(\mathbf{s}_n | \bm{\mu}, \bm{\lambda}, \bm{\pi}, \mathbf{A}, \bm{I})
    = \sum _{\mathbf{S}_{\backslash n}} p(\mathbf{S} | \bm{\mu}, \bm{\lambda}, \bm{\pi}, \mathbf{A}, \bm{I}),
    \end{eqnarray} 
    where we drop the upper suffix $(i)$ or $(i-1)$ on the iteration step for simplicity. Also, ${\mathbf{S}_{\backslash n}}$ is a subset of $\mathbf{S}$ minus $\mathbf{s}_n$. $p(\mathbf{s}_n| \bm{\mu}, \bm{\lambda}, \bm{\pi}, \mathbf{A}, \bm{I})$ is described with a following category distribution:
    \begin{eqnarray}
      p(\mathbf{s}_n| \bm{\mu}, \bm{\lambda}, \bm{\pi}, \mathbf{A}, \bm{I})
      =\mathrm{Cat}(\mathbf{s}_n|\bm{\eta}_n) 
      =\prod _{k=1} ^K \eta_{kn} ^{s_{kn}} 
      \label{cat}
      \label{psn} 
    \end{eqnarray}
      with 
      \begin{eqnarray}
      \eta_{kn}=\frac{\Tilde{\eta}_{kn}}{\sum _{k'=1} ^{K} \Tilde{\eta}_{k'n}} 
      \end{eqnarray}
      and 
      \begin{eqnarray} 
      \tilde{\eta}_{kn} = f_{kn} b_{kn}.
      \end{eqnarray}
      Here, $f_{kn}$ and $b_{kn}$ in the right-hand side are given as follows: 
      \begin{eqnarray}
      f_{kn}=\frac{\hat{f}_{kn}}{\sum _{k'=1} ^{K} \hat{f}_{k'n}}
      \end{eqnarray}
      with 
      \begin{eqnarray}
      \hat{f}_{kn}\!=\!\begin{cases}
      p(I_1|\mu_k, \lambda_k) \pi_k, &\!\!\!\!(n=1) \vspace{0.3cm} \\ 
      p(I_n|\mu_k, \lambda_k) \sum_{k'=1} ^{K} A_{kk'} f_{k'n-1} &\!\!\!\!(n\ne1)
      \end{cases} 
      \end{eqnarray}
      and 
      \begin{eqnarray}
      b_{kn}=\frac{\hat{b}_{kn}}{\sum _{k'=1} ^{K} \hat{b}_{k'n}}
      \end{eqnarray}
      with 
      \begin{eqnarray} 
      \hat{b}_{kn}\!=\!\begin{cases} 
      \sum_{k'=1}^{K} p(I_{n+1}| \mu_{k'}, \lambda_{k'}) A_{k'k} b_{k'n+1}, &\!\!\!\!(n\ne N) \vspace{0.3cm} \\
      1 &\!\!\!\!(n=N).
      \end{cases} 
      \end{eqnarray}
      In this calculation, $\bm{\mu}^{(0)}$, $\bm{\lambda}^{(0)}$, $\bm{\pi}^{(0)}$, $\mathbf{A}^{(0)}$ are required to find $p(\mathbf{s}_n| \bm{\mu}, \bm{\lambda}, \bm{\pi}, \mathbf{A}, \bm{I})$ [Eq.~(\ref{psn})] in the initial step. These are evaluated with the hyperparameters in TABLE~\ref{table-HMM}.

\item $p(\bm{\mu}, \bm{\lambda}|\mathbf{S}, \bm{\pi}, \mathbf{A}, \bm{I})$ [Eq.~(\ref{gibbs_mu_lambda})]: 
   
   This conditional probability distribution is described as the Gaussian-Gamma distribution: 
   \begin{eqnarray}
    & & p(\bm{\mu}, \bm{\lambda}|\mathbf{S}, \bm{\pi}, \mathbf{A}, \bm{I}) \nonumber \\
    &=& p(\bm{\mu}, \bm{\lambda}|\mathbf{S}, \bm{I}) \nonumber \\
    &=& \prod _{k=1}^K {\rm NG}(\mu_k,\lambda_k|\hat{m}_k,\hat{\nu}_k,\hat{a}_k,\hat{b}_k) 
    \label{gauss_gamma}
   \end{eqnarray}
   with 
   \begin{eqnarray}
     \hat{\nu}_k &=& \sum_{n=1}^N s_{kn}+\nu , \label{gibbs_nu_hat} \\
     \hat{m}_k &=& \frac{\sum_{n=1}^N s_{kn} I_n +\nu m}{\hat{\nu}_k}, \label{gibbs_m_hat} \\
     \hat{a}_k &=& \frac{1}{2} \sum_{n=1}^N s_{kn}+a, \label{gibbs_a_hat} \\
     \hat{b}_k &=& \frac{1}{2} \Biggl (\sum_{n=1}^N s_{kn} I_n^2+\nu m^2 -\hat{\nu}_k \hat{m}_k^2 \Biggr)+b,
     \label{gibbs_b_hat}
   \end{eqnarray}
   where, the original hyperparameters ($\nu, m, a, b$) of the Gaussian-gamma distribution are renormalized in as ($\hat{\nu}_k, \hat{m}_k, \hat{a}_k, \hat{b}_k$).

\item $p(\bm{\pi}| \mathbf{S}, \bm{\mu}, \bm{\lambda}, \mathbf{A}, \bm{I})$ [Eq.~(\ref{gibbs_pi})]: 
  
  This is the Dirichlet distribution: 
  \begin{eqnarray}
    p(\bm{\pi}|\mathbf{S},\bm{\mu},\bm{\lambda},\mathbf{A}, \bm{I}) 
    = p(\bm{\pi}|\mathbf{s}_1)
    = \mathrm{Dir}(\bm{\pi}|\bm{\hat{\alpha}}), 
  \end{eqnarray}
  where
  \begin{eqnarray}
  \hat{\alpha}_k=s_{k1}+\alpha_k \label{gibbs_alpha_hat}
  \end{eqnarray}
  with $\hat{\alpha}_k$ being the renormalized hyperparameter of the Dirichlet distribution. 

\item $p(\mathbf{A}|\mathbf{S}, \bm{\mu}, \bm{\lambda}, \bm{\pi}, \bm{I})$ [Eq.~(\ref{gibbs_A})]:

  This is also the Dirichlet distribution as 
  \begin{eqnarray}
  p(\mathbf{A}| \mathbf{S}, \bm{\mu}, \bm{\lambda}, \bm{\pi}, \bm{I})
  =p(\mathbf{A}|\mathbf{S})
  =\prod_{k'=1}^K \mathrm{Dir}(\mathbf{A}_{k'}|\bm{\hat{\beta}}_{k'}), \nonumber \\ 
  \end{eqnarray}
  where $\hat{\beta}_{kk'}$ is the renormalized hyperparameter from a bare parameter $\beta_{kk'}$ as  
  \begin{eqnarray}
  \hat{\beta}_{kk'}= \sum_{n=2}^N s_{k'n-1}s_{kn}+\beta_{kk'}. \label{gibbs_beta_hat}
  \end{eqnarray}
\end{enumerate}


%

\end{document}